\documentclass[structabstract]{aa}  
\usepackage{graphicx}
\usepackage[]{natbib}
\usepackage{rotating}
\usepackage{txfonts}
\newcommand{\gaia}{{\it Gaia }}
\newcommand{\gsp}{{\it GSP-Spec}}
\newcommand{\gspphot}{{\it GSP-Phot}}
\newcommand{\alfa}{$\alpha$}
\newcommand{\alfaFe}{[$\alpha$/Fe]}
\newcommand{\meta}{[M/H]}
\newcommand{\T}{$T_{\rm eff}$}
\newcommand{\g}{log($g$)}
\newcommand{\gunits}{cm/s$^2$}
\newcommand{\GRVS}{$G_{\rm RVS}$}

\begin{document}
   \title{Stellar parametrization from \gaia RVS spectra}

   \author{A. Recio-Blanco\inst{1}
          \and
          P. de Laverny\inst{1}
          \and
          C. Allende Prieto\inst{2,3}
          \and
          D. Fustes\inst{4}
          \and
          M. Manteiga\inst{5}
          \and
          B. Arcay\inst{4}
          \and
          A. Bijaoui\inst{1}
          \and
          C. Dafonte\inst{4}
          \and
          C. Ordenovic\inst{1}
          \and
          D. Ordo\~nez Blanco\inst{6}
          }

   \institute{Laboratoire Lagrange (UMR7293), Universit\'e de Nice Sophia Antipolis, CNRS, Observatoire de la C\^ote d'Azur, CS 34229, F-06304 Nice, France\\
              \email{arecio@oca.eu} 
         \and
         Instituto de Astrof\' isica de Canarias, E-38205 La Laguna, Tenerife, Spain
         \and 
    Universidad de La Laguna, Dept. de Astrof\'{\i}sica,
        E-38206 La Laguna, Tenerife, Spain
         \and 
         Departamento de Tecnolog\'ias de la Informaci\'on y de las Comunicaciones, Universidade da Coru\~na, E-15071, A Coru\~na, Spain
         \and
         Departamento de Ciencias de la Navegaci\'on y de la Tierra, Universidade da Coru\~na, E-15011, A Coru\~na, Spain
         \and 
         Geneva Observatory, University of Geneva, 51 ch. des Maillettes, 1290 Versoix, Switzerland
             }
 
   \date{Received September 2014; accepted 2014}

 
  \abstract
   {Among the myriad of data collected by the ESA
     \gaia satellite, about 150 million spectra will be delivered by the Radial Velocity
     Spectrometer (RVS) for stars as faint as \GRVS $\sim 16$. A specific
     stellar parametrization will be performed for most of these RVS spectra,
     i.e. those with enough high signal-to-noise ratio that should
     correspond to single stars having a magnitude in the RVS band
     brighter than $\sim$14.5. Some individual chemical abundances will
     also be estimated for the brightest targets.}
   {We describe the different parametrization
     codes that have been specifically developed or adapted for
     RVS spectra within the \gsp \ working group of the analysis consortium.
     The tested codes are based on optimization (FERRE and
     GAUGUIN), projection (MATISSE) or pattern recognition methods
     (Artificial Neural Networks).  We present and discuss their
     expected performances in the recovered stellar atmospheric
     parameters (effective temperature, surface gravity, overall
     metallicity) for B- to K- type stars.  The performances for the
     determinations of \alfaFe \ ratios are also presented for cool  stars.}
   {The different codes have been homogeneously tested with a
     large grid of RVS simulated synthetic spectra of BAFGK-spectral
     types (dwarfs and giants) with metallicities varying from
     $10^{-2.5}$ to $10^{+0.5}$ the solar metallicity, and considering
     variations of $\pm 0.4$~dex in the composition of the \alfa
     -elements.  The tests have been performed for
     S/N ratios ranging from 10 to 350}
   {For all the considered stellar types, stars brighter than
     \GRVS $\sim$ 12.5 will be very efficiently parametrized by the \gsp
     \ pipeline, including solid estimations of \alfaFe . 
     Typical internal errors for FGK metal-rich and metal-intermediate 
     stars are around 40~K in \T , 0.10~dex in \g ,
     0.04~dex in \meta , and 0.03~dex in \alfaFe \ at \GRVS=10.3. They degrade to
     155~K in \T , 0.15~dex in \g ,  0.10~dex in \meta , and 0.1~dex in \alfaFe \ 
     at \GRVS$\sim$12. Similar accuracies in \T \ and \meta \ are found for A-type stars, 
     while the \g \ derivation is more accurate (errors of 0.07 and 0.12~dex
     at \GRVS=12.6 and 13.4, respectively). For the faintest stars, with \GRVS$\ga$13-14,
      a \T \ input from the spectrophotometric derived parameters  will allow the 
    improvement of the final \gsp \ parametrization. }
  {The reported results, although neglecting possible 
mismatches between synthetic and real spectra, show that the contribution of the RVS based stellar parameters will be unique
    in the brighter part of the \gaia survey allowing crucial age estimations, and accurate
    chemical abundances.  This will
    constitute a unique and precious sample for which many pieces of
    the Milky Way history puzzle will be available, with unprecedented
    precision and statistical relevance.}


   \maketitle
%

\section{Introduction}
\label{Intro}

The European Space Agency \gaia \ mission was successfully
launched on December 19, 2013 from the French Guiana Space centre in
Kourou.  Gaia is an ambitious astrometric, photometric and
spectroscopic survey of a large part of the Milky Way: about 1\% of
the Galactic stellar content down to $V  \sim 20^{th}$ magnitude will be
observed several tens of times.  \gaia will thus revolutionize our
view of the Galaxy together with our understanding of its formation
and evolution history. Apart from the astrometric instrument, two
low-resolution spectrophotometers (Blue and Red Photometers, BP/RP)
and the Radial Velocity Spectrometer (RVS) are in operation onboard of
\gaia .  Recent summaries of these instrument
characteristics, their main goals, and expected scientific impact can
be found in \citet{deBruijne12,BailerJones13}\footnote{see also http://www.cosmos.esa.int/web/gaia/science-performance}.

The analysis of the \gaia data is performed by the {\it Data
Processing and Analysis Consortium} (DPAC) composed of nine {\it
Coordination Units} (CU). One of them, CU8 {\it Astrophysical
Parameters}, is in charge of the classification and the
parametrization of the observed targets \citep[see][]{BailerJones13}.
The CU8/{\it Working Group} that is responsible for the
main parametrization of the RVS spectra is the {\it Global Stellar
Parametrizer - Spectroscopy} (\gsp ). 
To complement this \gsp \ parametrization, other CU8 modules ({\it Extended Stellar Parametrizer, ESP})
will also estimate atmospheric parameters from RVS spectra of more specific types of stars as
emission-line stars ({\it ESP-ELS}), hot stars ({\it ESP-HS}), cool stars ({\it ESP-CS}) 
and, ultra-cool stars ({\it ESP-UCD}) \citep[see][]{BailerJones13}.
\gsp \ is composed of three
research groups having different and complementary expertises in
automated stellar classification from spectral data. In this article,
we present the parametrization performed within this working group for
RVS spectra.

The RVS is a high-resolution integral-field spectrograph that will
survey the whole sky at a rate of about 100 spectra per second, producing about
15~billion of spectra during the mission. From its preliminary real 
performances, the RVS will collect spectra with large enough
signal-to-noise ratio (S/N) in order to derive their radial velocity
for stars brighter than \GRVS $\la 16$ (i.e. about 150 million stars,
\GRVS \ being the \gaia magnitude of the targets through the RVS
filter). This limiting magnitude corresponds to $V \la 17.3$ for
a solar-type star (for the corresponding magnitudes in other photometric Gaia bands, see 
Tab.~\ref{Tab_GRVS} and Fig.~\ref{FigColorColor} presented in Sect.\ref{GridRandom}).
Several tens of millions of stars will be observed by the RVS down to a magnitude of
 \GRVS $\la 13$, and about 5 million stars down to \GRVS $\la 12$.

The RVS will provide spectra in the CaII IR triplet region
(from 847 to 871~nm\footnote{The RVS red edge has been
shifted from 874 to 871~nm with respect to the original configuration, following a change in the RVS filter
effective transmission. In the present work, we have thus adopted this 871~nm cut-off for
the simulated spectra.})  at  a spectral resolution of $\sim$11\,200\footnote{The design specification
of R$\sim$10\,500 being exceeded.}. In addition to the CaII strong lines,
one encounters, in the RVS spectral range and for late-type star spectra,
weak lines of Si~I, Ti~I, Fe~I, etc. In hotter (A-F types) star spectra, weak lines of N~I,
S~I, and Si~I appear around the Ca~II lines and the strong Paschen hydrogen
lines (see Fig.~\ref{Spec1} \& \ref{Spec2}).  Even hotter (\T $\ga$
15\,000~K) stellar spectra contain lines of N~I, Al~II, Ne~I, and
Fe~II whereas the Ca~II lines start to decrease, and some He~I lines
appear.

 On the other hand, the \gaia commissioning phase has revealed that 
the RVS suffers from (i) a level of scattered light higher than expected, and variable with time and 
CCD position (mainly sunlight scattered around sunshields) and thus an increased noise
for part of the spectra, together with (ii) a time-variable throughput loss due to mirror contamination that reduces
the collected signal by a few tenths of a magnitude. This last issue is regularly corrected 
thanks to de-contamination campaigns that reduce the loss to acceptable levels.
Moreover, the \gaia DPAC has put in place a new 
version of the on-board software which results in a data collection scheme which is more robust against the stray light. This is mainly realised by lowering the auto-collimation width of read-out windows in the RVS (and possibly the astrometric field) such that less noise due to stray light is accumulated. The new video processing unit software that makes this possible has already been uploaded to the satellite \citep{Fleitas2015}.
In addition, following the actual RVS performances revealed by the commissioning phase,
it has also been decided that every RVS spectra will be provided in the nominal high-resolution
mode to minimize the background contamination that is a function of the window width.  
Initially, a binning by a factor three for stars fainter than \GRVS$<$10 was planned \citep[e.g.][]{BailerJones13}, 
decreasing the effective resolution to around 7500. This possibility
has now been definitely abandoned. 

In this paper, the above mentioned post-launch
RVS characteristics are taken into account and the new S/N-magnitude relation recently published by the European
Space Agency has been adopted\footnote{See http://www.cosmos.esa.int/web/gaia/sn-rvs}.
In addition, the expected final \gsp \ performances are given for the actual RVS resolution of $\sim$11\,200. Nevertheless,
the influence of the effective resolution change, at a constant S/N value, is studied in Sect.~\ref{Res} .

The main goal of the RVS is to measure the radial velocity of the
stars in order to get their full 3D space motions when combined with
the \gaia proper motions.  However, the RVS data will be also very
useful to parametrize brighter stars observed by \gaia in
complement to the parametrization performed independently from the two
more sensitive low-resolution spectrophotometers \citep[see][for an
estimate of the expected parametrization performances with BP/RP
data together with the performance improvements presented in \citet{BailerJones13}
and Andrae et al. (2015, in preparation)]{Liu12}.  
We point out that, in the present work, we do not
consider for \gsp \ any ({\it a-priori}) input from the BP/RP
parametrization although this is one of the alternatives implemented
in the \gaia global data processing pipeline developed by CU8 and called 
Astrophysical parameters inference system
\citep[Apsis, see][]{BailerJones13}.

Since \gaia scans the whole sky, each target will be observed several
tens of times depending on their location on the sky (with an average
of 40 epochs per stars for the RVS at the end of the mission, 
assuming the nominal 5-year mission).  As a
consequence, the S/N of the combined spectra will increase with time
during the mission and we will have to re-parametrize these better
quality spectra delivered by the successive releases.  In the
following, we consider RVS end-of-mission spectra that are a combination of the
successive individual observations.
It is expected that any star brighter than \GRVS $\la$~14 (i.e. several tens of millions)
 will be parametrized by \gsp . 
The estimated stellar parameters will be the effective temperature (\T ), the surface
gravity (\g ), the global metallicity (\meta ), and the abundance of \alfa -elements
versus iron (\alfaFe ). In a second step and whenever possible (depending on
spectral type, metallicity, S/N ratio, radial velovity shifts...), the individual chemical abundances
of several species as Fe, Ca, Mg, Ti, and Si will be also measured. This should
be performed for about five million sources with
an expected accuracy of about 0.1~dex for \GRVS $\la$ 12 
owing to a specific optimization method 
that is under development \citep{Guiglion14}.

The amount of spectra to parametrize
requires the use of automated methods that have to be fast enough and able to
manage different types of stars.  Therefore, \gsp \ is currently
composed of different independent codes based on different algorithms.  The advantage of having
different codes is to get independent parameter estimates and to
get the best parameters all across the studied space (a given
algorithm could provide excellent results for some parameter
combinations and/or S/N ratios but rather poor results for others). Therefore, every RVS
spectra will be parametrized by the different \gsp \ codes and
most stars will be assigned to several parameter estimates with
quality flags.

In this paper, we describe the \gsp \ codes and their expected
performances for different types of stellar populations as they have
been implemented at the \gaia launch epoch. 
 Increasingly optimized versions of the \gsp \ module of Apsis \citep[see][]{BailerJones13}
 are delivered at each operations cycle. \gsp \ is expected to 
 be running in operations cycle~4 in 2017, with a possible contribution from
 the third \gaia \ data release (planned around 2018).

 In the following, we
first present in Sect.~\ref{Algo} the codes specifically
developed or adapted to the RVS stellar spectra by \gsp \ in order to
estimate their stellar atmospheric parameters together with their
enrichment in \alfa -elements with respect to iron.  We then detail
in Sect.~\ref{Tests} the methodology adopted for testing and comparing
these different codes and their \gsp \ performances on simulated spectra
at different RVS magnitudes are investigated and discussed in
Sect.~\ref{Perf}. 
The end-of-mission \gsp \ expected parametrization is described in 
Sect.~\ref{final} for different types of stars and S/N ratios.
We then provide 
a comparison with the expected
performances for the parametrization from BP/RP photometry
(Sect.~\ref{GSPPhot}) and from some ground-based spectroscopic surveys (Sect.~\ref{Surveys}).
We finally conclude this work in Sect.~\ref{Conclu}.


\section{The \gsp \ automated parametrization codes}
\label{Algo}
In this work, four different codes for estimating the atmospheric stellar
parameters and the overall \alfaFe \ chemical index (the latter, for cool FGK-spectral
type stars, only) have been tested for \gsp
: FERRE, GAUGUIN, MATISSE, and Artificial Neural Networks (ANN). These
codes have already been extensively described in previous
papers. We hereafter refer to the indicated references for their
detailed description. We below provide only a brief summary of these
codes and we mainly focus on the specificities that have been
developed for \gsp . Furthermore, we point out that most of these
codes have already been separately used to parametrize real
stellar spectra collected at a similar resolution and over the same
wavelength region as the ones of the RVS (see references below).
However, the present work is the first one to illustrate and compare
their performances in the \gaia context.

Within the different data mining approaches that have been developed
so far, \gsp \ parametrization algorithms belong to the class using
the mapping with reference models in a continuously variable space. In
addition, they belong to the three main families of parametrization:
optimization methods, projection methods and pattern recognition.

\begin{itemize}
\item The optimization codes (FERRE and GAUGUIN) perform a
distance minimization between the full input spectrum and the reference
spectra grid.

FERRE\footnote{The code is now public and available at: http://hebe.as.utexas.edu/ferre} \citep{Allende06} is a FORTRAN90 code that
compares the $\chi^2$ between models and observations to identify
the optimal set of atmospheric parameters (and/or abundances) of a
star. The search is performed using the Nelder-Mead (1965)
algorithm. The model evaluation is based on linear interpolation in a
pre-computed grid, which is held on the computer's memory for speed.
The observed and (interpolated) model spectra are forced to have the
same mean value. Multiple searches are performed for each spectrum,
initialized at 100 random positions in the grid. The code is parallelized over
multiple cores using OpenMP and is made available upon request.

GAUGUIN \citep{Bijaoui12} is a classical local optimization method
implementing a Gauss-Newton algorithm. It is based on a local
linearisation around a given set of parameters that are associated to
a reference synthetic spectrum (via linear interpolation of the
derivatives). A few iterations are carried out through linearisation
around the new solutions, until the algorithm converges towards the
minimum distance.  In practice, and in order to avoid trapping into
secondary minima, GAUGUIN is initialized by parameters independently
determined by other algorithms such as MATISSE (this is noted hereafter
as MATISSE$_G$, see below).  GAUGUIN
is part of the analysis pipeline of the \gaia- ESO Survey
\citep[GES,][]{Gilmore12} for the GIRAFFE spectra (Recio-Blanco et
al., 2015, in preparation).

\item The MATISSE algorithm \citep{Recio-Blanco06} is a projection
method in the sense that the  full input spectra are projected into a set
of vectors derived during a learning phase, based on the
noise-free reference grids (see Sect.~\ref{Grids}). 
These vectors are a
linear combination of reference spectra and could be roughly viewed
as the derivatives of these spectra with respect to the different
stellar parameters.  
MATISSE is thus a local multi-linear regression
method.  Furthermore, a two-step procedure in the parameter
estimation has been implemented within MATISSE to tackle
non-linearity problems. 
Other recent applications of MATISSE to very large amounts of real
observed spectra are the {\it AMBRE} project \citep{deLaverny13,
Worley12}, the RAVE fourth Data Release \citep{Kordo13} and the
\gaia- ESO Survey (see Recio-Blanco et al., 2015, in preparation).

\item Finally, the ANN code uses a pattern recognition approach to 
parametrise the spectra.


The implementation of the ANN method for \gsp \ has already been
presented in \citet{Manteiga2010}.  Shortly, the architecture is a
feed-forward network with three layers, trained with the online error
back propagation algorithm.  We generate one network for each stellar
parameter where the number of neurons in the input layer coincides
with the number of points of the format that was selected as signal
representation.  The output layer consists of the parameter to be
predicted.  The activation function for the hidden and output neurons
of the network is a sigmoidal function.  We also set the number of
hidden neurons as the minimum between half the dimensionality of the
input signal and 200 hidden units at maximum to reduce the
computational burden of the training process.  To deal with the
initialization dependence and possible local minima, a series of
training procedures are performed, until a near-optimal value is
reached. On the other hand, early stopping is also used by means of
a validation dataset to avoid the problem of over fitting, so that the
network state which best generalizes is kept.  Furthermore, to
generalize the application to random spectra, we use 100 noised
reference grid spectra during the training phase (see below).  Finally,
we made use of the wavelet decomposition to obtain new signal representations 
(low pass filter) which are used as the network inputs, both in the
training and testing phases. In practice, this means that the results of this code
are obtained by adopting the approximations of the first, second or third
level in the wavelet pyramid (depending on the S/N value) as input for the ANN instead of the full spectra.
\end{itemize}

As outputs, these four codes  provide the three
atmospheric parameters (\T , \g , and \meta ) and the \alfaFe
\ chemical index (for FGK stars only) estimates together with their associated
uncertainties (except for the ANN).  For some of them (FERRE, GAUGUIN, MATISSE), 
a quality control parameter based on the goodness of the
fit between the input spectrum and an interpolated one at the
estimated parameters is also provided.
We report in Tab.\ref{Tab_codes}
more details on some technical aspects of these codes.

\begin{table}
\sidecaption
\caption{Summary of technical aspects of the tested codes or algorithms.}
\label{Tab_codes}
\centering 
\begin{tabular}{lccc}
& & & \\
\hline
& & & \\
& Training & Programming & Already implemented \\
& phase    & language    & in Apsis \\
& & & \\
\hline
& & & \\
FERRE   & No  & Fortran         & No \\
GAUGUIN & No  & Fortran \& Java & Yes \\
MATISSE & Yes & Fortran \& Java & Yes \\
ANN     & Yes & Java            & No \\
& & & \\
\hline
\end{tabular}
\end{table}


\section{Adopted methodology for quantifying the code performances}
\label{Tests}
We present here the implemented homogeneous tests, including the
used data and the analysis methodology, that estimate and
compare the expected performances of the different \gsp
\ codes. The ultimate goal is to determine the optimal
application fields of the different codes in terms of stellar
types and S/N.  This will lead to the definition of the quality flags
that will be assigned to the different parameter estimates
within the \gsp \ pipeline.

In this Section, the data used for the tests and the considered 
methodology for the analysis of the results are described.
The different subsections present the steps of the followed
procedure. In particular:

\begin{itemize}
\item  The codes have been trained (when necessary, depending
on the algorithm strategy, see Sect.~\ref{Algo}) using large grids of
noise-free RVS simulated synthetic spectra that are described in
Sect.~\ref{Grids}\footnote{We however point out that the non-trained
codes also use the spectra grid for their parametrization based on a 
fitting process.}

\item Then, in order to perform an homogeneous
comparison of the methods, parametrisation tests have been performed using
noised random grids. The six adopted S/N values are 350, 150, 125, 40,  20 and 10.
Those test grids contain interpolated spectra with arbitrary
parameter values (see Sect.~\ref{GridRandom}) that span the whole grid parameters
range.  

\item A subsample of the previously noised random spectra, excluding non-physical
parameter combinations and correctly populating the Hertzsprung-Russel diagram,
has been selected (c.f. Sect.~\ref{RandomSamples}). 

\item Finally, we have tackled the problem of the recently abandoned RVS spectra rebin
for stars with \GRVS$>10$ (c.f. Sect.~\ref{Res}). The influence of this change in
the \gsp \ performances is quantified thanks to specific tests with one of the tested
algorithms.

\end{itemize}

  We would like to point out that,  in this work, we
favoured the use of synthetic spectra instead of real ground based observed ones (RVS data
not being available yet) in order to
explore any possible combination of the four parameters over a wide range of
possible values and to keep a good homogeneity between the tests. We
are fully aware that synthetic spectra may not be perfectly realistic
when compared to observed stars for some parameter combinations but
this will not affect the codes comparison.  Of course, the
application to real observed spectra will lead to larger errors ({\it
external ones}) mostly due to the possible mismatches that could exist between synthetic and
observed spectra. These effects will be quantified (and possibly
corrected) during the mission owing to calibration with {\it reference
stars} that will be (or already are for some of them) accurately
parametrized \citep[see][]{BailerJones13}.

Finally, we neglect in the following any effects that could be induced
by wrong normalization or radial velocity corrections of the input
spectra. Within the \gaia analysis pipeline, such problems will be
examined, in collaboration
with \gsp, by the CU6 in charge of providing RVS spectra to CU8. 
The synthetic spectra described below are thus all at the rest frame
and normalized.  We, however, point out that, as the estimation
of the atmospheric parameters and individual abundances can be quite
sensitive to the pseudo-continuum normalization, an automated
renormalization of the input spectra has already been implemented
within \gsp . Indeed, the RVS spectra can be renormalized through an
iterative procedure coupled with the atmospheric parameters
estimates. Such an iterative procedure has already be shown to be very
successful when automatically applied to real spectra \citep[][and
the GES pipeline, Recio-Blanco et al., 2015, in
preparation]{Kordo11a, Worley12, Kordo13}.  
More specifically, we also point out that \cite{Kordo11a} have already shown that parametrization 
errors induced by normalization uncertainties as large as 3\% of the continuum level for RVS-like
spectra can be neglected if such an iterative renormalisation is implemented.
This is however not
considered in the following tests since we remind that we focus on already
perfectly normalized input synthetic spectra.

\subsection{Grids of reference spectra}
\label{Grids} 

\begin{figure*}[ht!]
\sidecaption
\includegraphics[height=12.cm,width=14.cm]{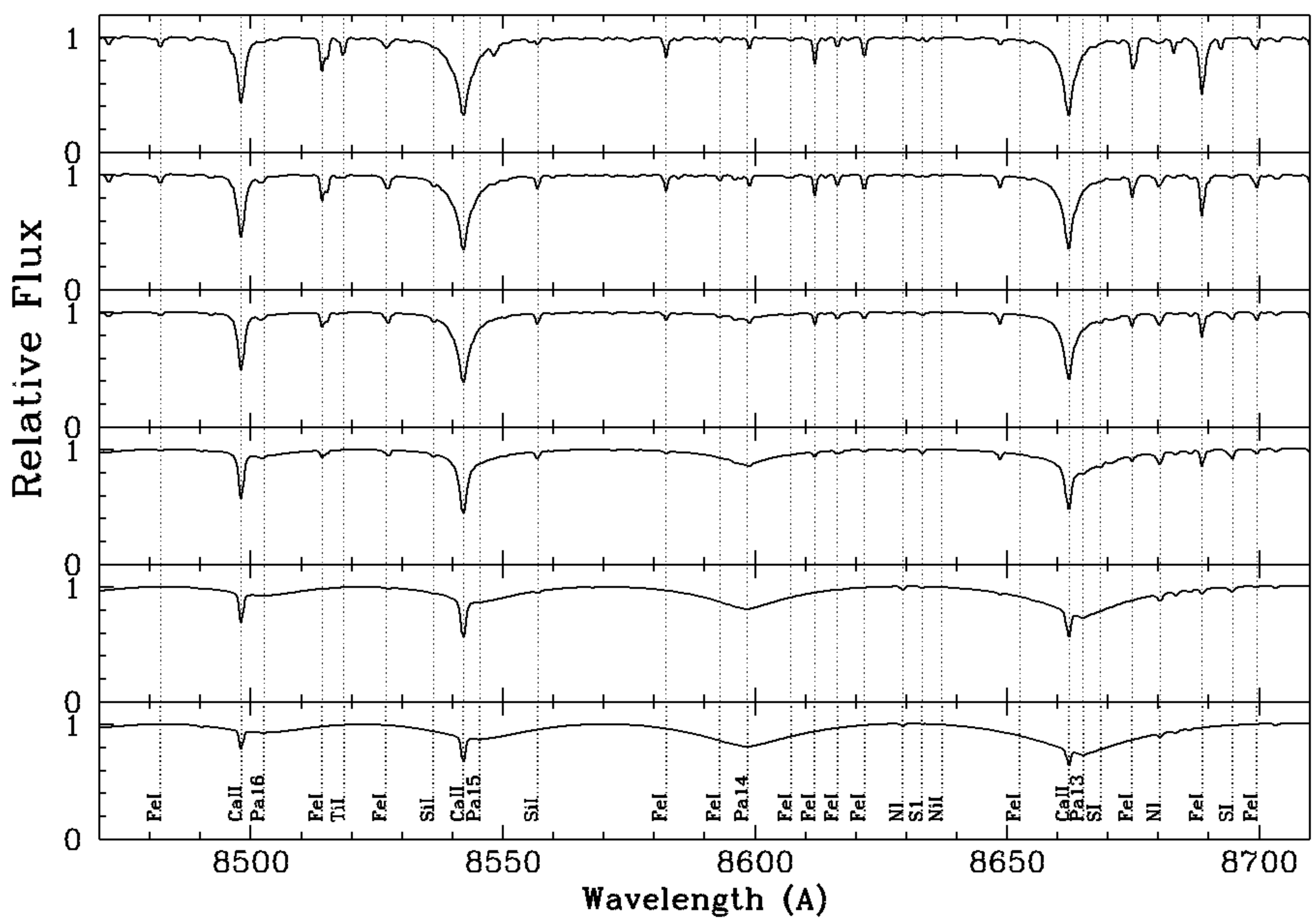}
\caption{Noise-free simulated RVS spectra for late B to K-spectral type metal-rich dwarf
  stars, included in the {\it reference} grids.
  The adopted effective temperatures are 11\,000~K,
  9\,000~K, 7\,500~K, 6\,500~K, 5\,500~K, and 4\,500~K from bottom to
  top, respectively. The other stellar parameters are kept constant at
  \g = 4.5~\gunits , \meta = +0.0~dex, and \alfaFe = +0.0~dex.  The identified lines
  refer to most of the strongest atomic transitions that are present in the
  different spectra. }
\label{Spec1}
\end{figure*}

\begin{figure*}[ht!]
\sidecaption
\includegraphics[height=12.cm,width=14.cm]{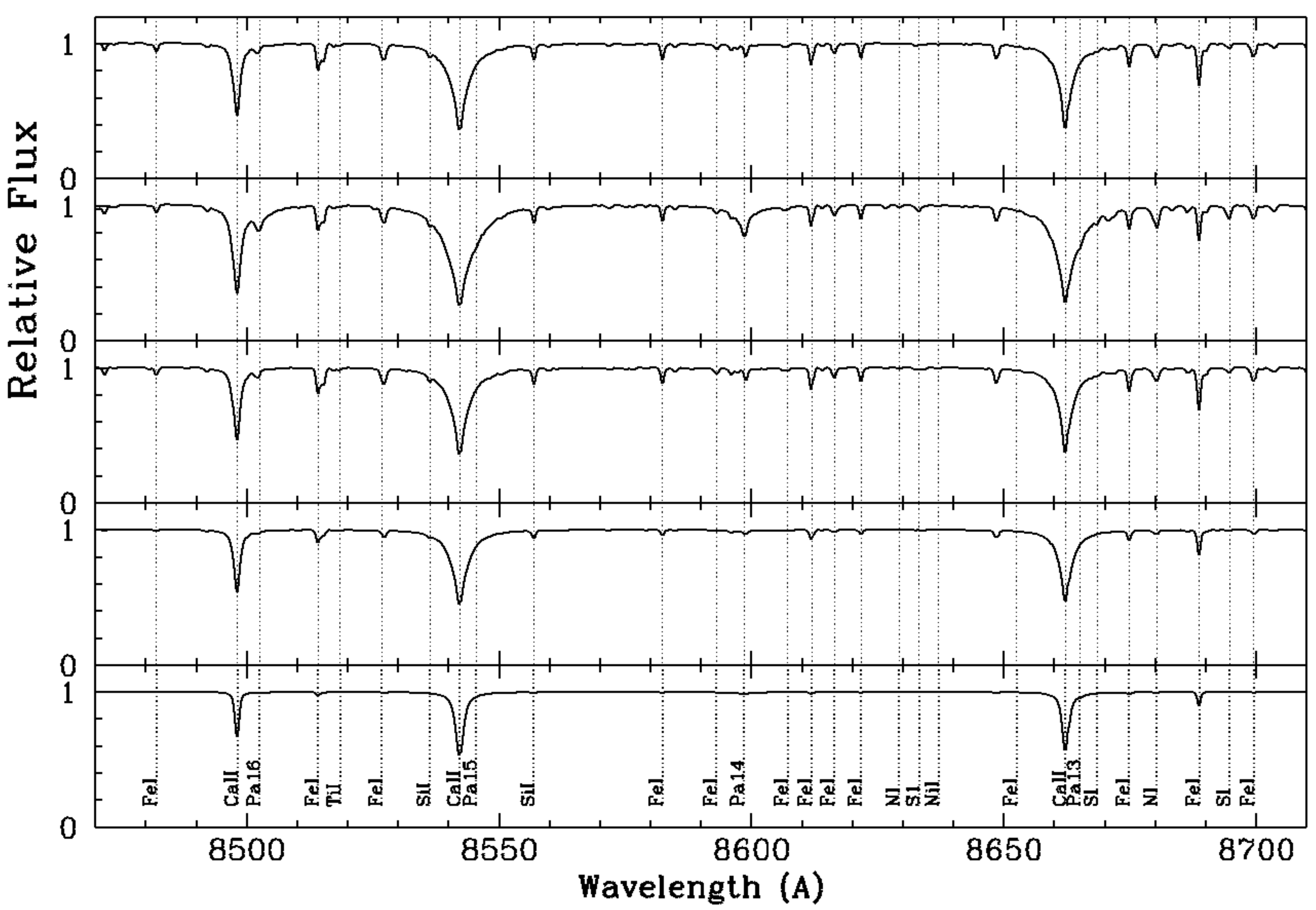}
\caption{Same as Fig.~\ref{Spec1} but for cool stars only. Taking as reference
  a Solar spectrum defined with 5\,750~K, \g = 4.5~\gunits , \meta = +0.0~dex,
  \alfaFe = +0.0~dex and R$\sim$11\,000 (top panel), the following panels 
  show the effect of (from top to bottom):
  a change in surface gravity to \g  = 2.0~\gunits \ (second panel);
  and changes in the chemical composition (third, fourth and bottom panels 
  showing respectively \meta = +0.5~dex and \alfaFe = +0.2~dex; \meta = -1.0~dex and 
  \alfaFe = +0.4~dex; \meta =  -2.0~dex and \alfaFe = +0.4~dex).}
\label{Spec2}
\end{figure*}

\begin{figure*}[t!]
\includegraphics[height=8cm, width=18.cm]{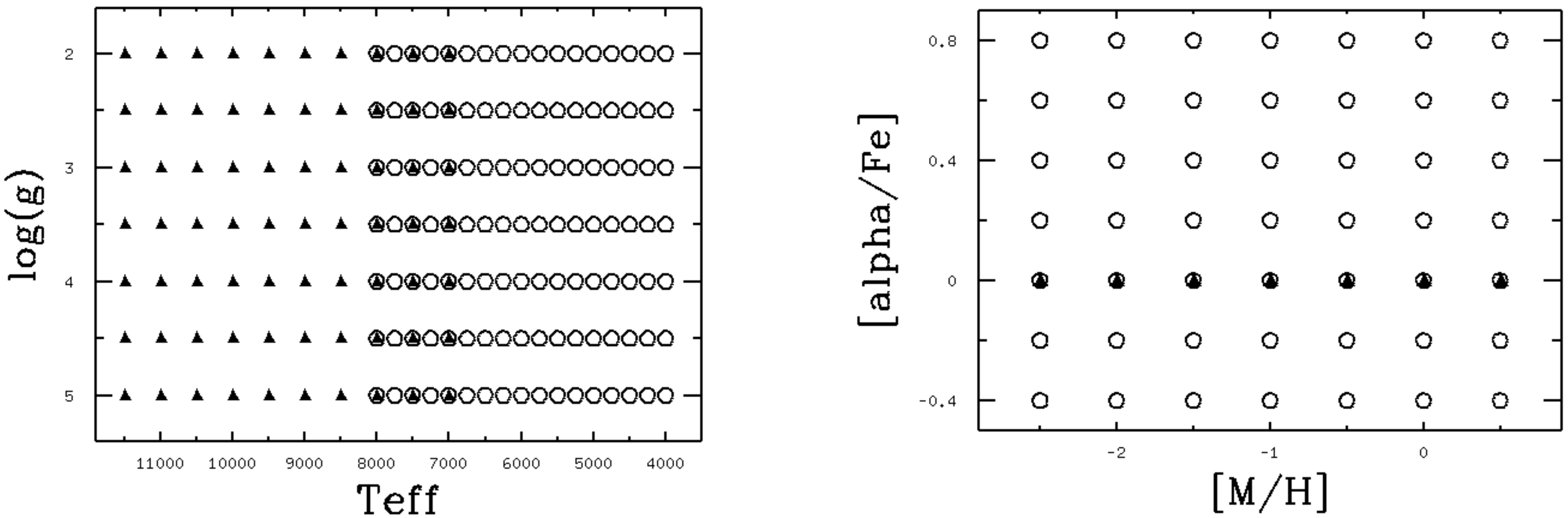}
\caption{Distribution of the {\it reference} synthetic spectra in the
  atmospheric parameter space (\T , \g , \meta \ and, \alfaFe \ are in K, 
  \gunits , dex and, dex, respectively).
  The cool-star and hot-star grids are
  shown with open circles and filled triangles, respectively.  Any
  combination of the shown stellar parameters has been considered when
  building these grids, except for the hot-grid in which no variations
  in \alfaFe \ have been considered (see filled triangles in the right
  panel).}
\label{FigGridReference}
\end{figure*}

The reference grids are a collection of noise-free high-resolution
normalized synthetic spectra that have been computed from Kurucz model
atmospheres.  We favoured these Kurucz models, contrary to previous works
based on MARCS model atmospheres \citep[see, for
instance,][]{deLaverny12}, because they allow us to consider
consistent spectra for cool FGK and hot BA-spectral type stars.  
The original computed spectra preserved
    the continuum slope, but then we continuum normalized each
    spectrum by (i) fitting iteratively a line and $\sigma$-clipping 10 times
    points that fell between 0.1$\sigma$ and 3$\sigma$ below and
    above the fit, respectively, and (ii) dividing the original spectrum
    by this final fit. We refer in the following to such {\it normalized} spectra,
in which the slope in the spectrum, within the RVS wavelength domain, is not
conserved. 
In practice, thanks to the estimates of stellar distance,
and after accurate flux calibration, it would also be possible to
analyse \gaia/RVS absolute flux spectra. This possibility is however
not considered in this work.

Regarding the grid calculations, they were computed
    using \citet{Castelli03} model atmospheres and a
    line list from Kurucz\footnote{from his website kurucz.harvard.edu}
    enhanced with damping constants
    from \citet{Barklem00} for atomic transitions. The calculations
    were done using the ASSET synthesis code \citep{Koesterke08, Koesterke09},
    sampling the spectra with steps corresponding to 1~km/s.
    The values of the solar reference abundances in the synthesis
    are from \citet{Asplund05}, while those of the corresponding
    atmospheric structures are from \citet{Grevesse98}. The
    abundances of the \alfa -elements were changed for the synthesis,
    but not in the model atmospheres.

Four parameters were considered for building the FGK-spectral type 
grid. These parameters are the effective temperature \T \ that varies
from 4\,000 to 8\,000~K (with a step of 250~K), the surface gravity \g
\ varying between 2.0 to 5.0~\gunits \ (step of 0.5~dex), the mean
metallicities varying from $10^{-2.5}$ to $10^{+0.5}$ the solar
metallicity (with a step of 0.5~dex), and variations of $\pm 0.4$~dex
in the enrichment in the \alfa -chemical species with respect to iron
(step of 0.2~dex).  This cool-star reference grid contains 5\,831
spectra.

For hotter stars (from late B to F-spectral types), only the first
three parameters were considered since the spectra become almost metal
line-free with increasing \T . The effective temperature for this
hot-star grid varies from 7\,000 to 11\,500~K (step of 500~K). The
surface gravity and mean stellar metallicity ranges together with their variation
steps are identical to those of the cool-star grid (without variations
in \alfaFe ). We end up with a reference grid of 490 hot stellar
spectra.

We point out that any possible combination of the above mentioned
stellar parameters has been considered to build these {\it square}
cool and hot grids (i.e. without gaps in any
 of these three or four parameters).

In addition, these very high-resolution grids have then been degraded in
order to mimic the RVS instrumental effects by convolution with a
Gaussian profile for the spectral resolution (R~$\sim$11\,200) and
adopting a sampling of $\sim$0.024~nm/pixel (1\,125~pixels in total).
Additionally, a sampling of $\sim$0.072~nm/pix (375~pixels) has been
considered to produce two low resolution grids (R~$\sim$7\,500). They have been used to
analyse the influence of the post-launch effective resolution change (c.f. Sect.~\ref{Res}).

In the following, we will refer to these two RVS synthetic spectra
grids as the {\it reference} grids.  For illustration, some spectra
corresponding to the late B- to K-spectral types are shown in
Fig.~\ref{Spec1} together with the identification of some of their strongest
lines. Moreover, Fig.~\ref{Spec2} shows some cool stars spectra
representative of different Galactic populations.  Finally, the
distribution of the atmospheric parameters and the \alfaFe \ chemical
index of the reference grids is shown in Fig.~\ref{FigGridReference}.

\subsection{Grid of noised random spectra}
\label{GridRandom}
  From the {\it reference} high-resolution (R~$\sim$11\,200) and low-resolution (R~$\sim$7\,500) noise-free grids described above, 
interpolated noised synthetic spectra have been generated for the
codes tests.

The interpolations were performed at random combinations of the four
parameters.  The S/N has then been
simulated by adding noise (see below) to these interpolated spectra.
The six adopted S/N values are 350, 150, 125, 40,  20 and 10.
They correspond to the
RVS filter magnitudes \GRVS \ values equal to 
8.4, 10, 10.3, 11.8, 12.6, 13.4 according to the most recent RVS performances, including post-launch studied instrumental
effects like straylight contamination, actual line spread function profiles, effects of window decentering and light loss.

\begin{figure}[ht]
\includegraphics[height=4.4cm, width=8.4cm]{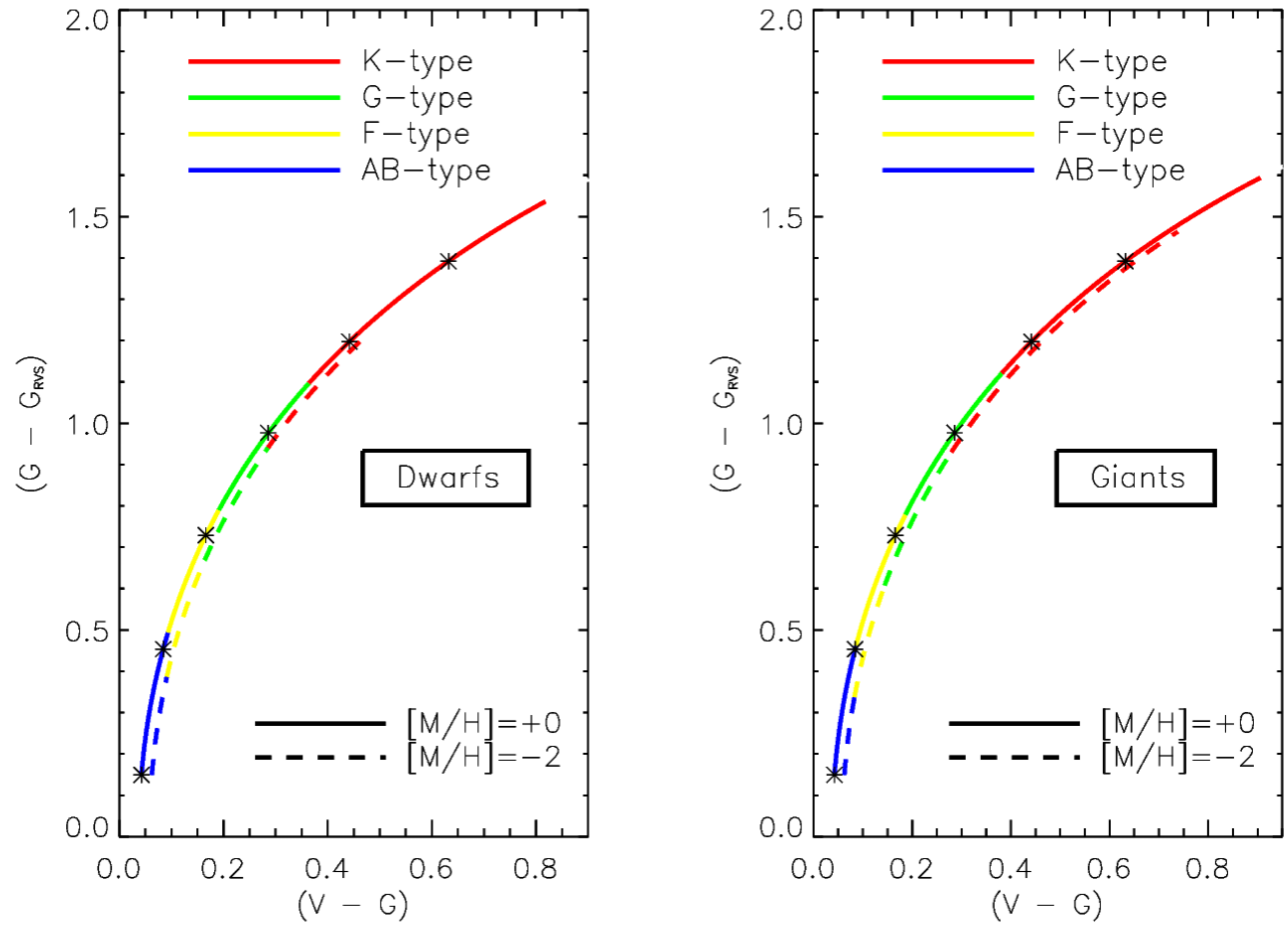}
\caption{($G$-\GRVS ) versus ($V$-$G$) relation for the stellar types (colour-coded) defined in Sect.~\ref{final}.
For clarity reasons, the dotted lines (metal-poor stars) have been slightly (+0.02) horizontally shifted and the dwarf and giant regimes separated. 
The ($G$-\GRVS ) and ($V$-$G$) indexes only vary with the ($B$-$V$) colour index but the covered ranges in both axis are dependent
on the metallicity.
The crosses along the curves refer to ($B$-$V$) varying from 0.0 to 1.5 (step of 0.25) from bottom to top.}
\label{FigColorColor}
\end{figure}

We again point out that this added noise will allow us to estimate the parameter
uncertainties caused by internal and instrumental errors and not the external errors, mainly dominated by possible synthetic spectra
mismatches.
Moreover, we  report in Tab.~\ref{Tab_GRVS} the magnitudes in the $G$ and $V$-bands corresponding to \GRVS = 13.5
for different stellar types and in Fig.~\ref{FigColorColor} the ($G$-\GRVS ) versus ($V$-$G$) relation for the range of \T \ studied in the
present work. These estimated $V$, $G$ and \GRVS -magnitudes and associated colours are based on 
the colour transformations provided by \cite{Jordi10} together with the photometric
colours -- \T \ relations of \cite{Ramirez05} and \cite{Boyajian12} for cool and hot stars, respectively.

\begin{table}
\sidecaption
\caption{$V$ and $G$-magnitudes corresponding to \GRVS = 13.5 for
some stellar types as defined in Sect.~\ref{final}.}
\label{Tab_GRVS}
\centering 
\begin{tabular}{lcc}
& &\\
\hline
& &\\
Stellar type &  $V$ (mag) & $G$ (mag)\\
& &\\
\hline
& &\\
B dwarf & 13.69 & 13.65\\
A dwarf & 13.83 & 13.77\\
& &\\
F metal-poor dwarf & 14.25 & 14.12 \\
F metal-rich dwarf & 14.32 & 14.18 \\
G metal-poor dwarf & 14.54 & 14.33 \\
G metal-rich dwarf & 14.76 & 14.48 \\
K metal-poor dwarf & 15.06 & 14.66 \\
K metal-rich dwarf & 15.37 & 14.82 \\
& &\\
G metal-poor giant & 14.54 & 14.33 \\
G metal-rich giant & 14.76 & 14.48 \\
K metal-poor giant & 15.14 & 14.70 \\
K metal-rich giant & 15.52 & 14.89 \\ 
& &\\
\hline
\end{tabular}
\end{table}

Practically speaking, we have simulated end-of-mission RVS spectra
based on the instrument performance information available to us.  The
CCD windows for the spectra, and therefore their  noise
properties depend mainly on the source brightness. On the other hand, we take
into account that the binning in the spectral direction corresponds to
one RVS pixel\footnote{Note that the RVS CCDs, as well as the other
  \gaia instruments, are operated in TDI mode, and hence talking about
  pixels (or {\it sample} as often used in \gaia \ literature)
  is not accurate, since the signal in any sample of a spectrum
  has been accumulated over many pixels as the star crosses the focal
  plane, but we will use this term for analogy.} 
  
The average signal per binned pixel in one RVS spectrum $I$ (single
visit and single CCD) is approximately defined by the brightness in
that band as \GRVS ~= -2.5*log($I$ x $q$) + 22.5866 where $q$, the
number of binned pixels, is equal to 1\,260 for the high resolution and $q$=420 for the low one.
We account for Poisson shot noise in the data, and the CCD
readout noise, assumed to be 4 e-.  
Since the final RVS spectra will be accumulated from a variable number
of visits, depending mainly on the location of a source on the sky,
and since typically objects cross three CCDs per visit, we simulate
individual observations (spectra acquired per CCD per visit), and then
combine them to produce an end-of-mission spectrum for each source.

 The final high and low resolution noised grids contain 20\,000 random spectra at each selected
\GRVS \ magnitude. They cover a wide range of atmospheric parameter
values, including some non-physical combinations (unreal stars).
Actually, 10\,000 random spectra have been interpolated in each
high/low resolution and cool/hot {\it reference} grids, having stellar parameters independent 
from the \GRVS \ values.  Moreover, other additionally 100
random noised spectra were selected to train independently the
ANN code (see Sect.~\ref{Algo}). These grids are made available 
in FITS format upon request to the authors.

\subsection{Final definition of noised random spectra samples}
\label{RandomSamples}

\begin{figure}[t!]
\includegraphics[height=8.4cm, width=8.4cm]{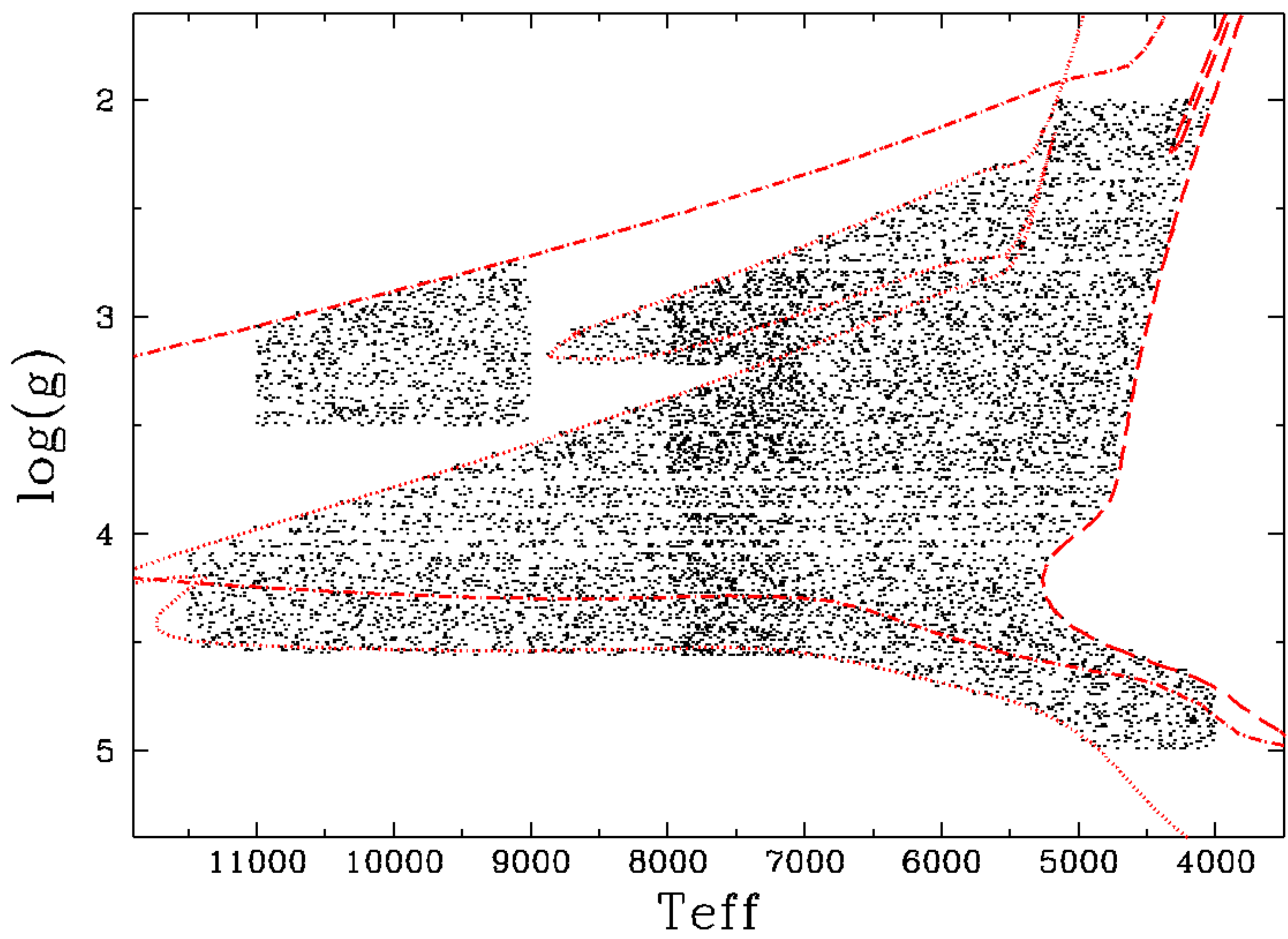}
\caption{Hertzsprung-Russell diagram of the random spectra (black
  dots) produced for the code performance tests (\T \ and \g \ are 
  in K and \gunits , respectively). The isochrones
  in long-dashed, dotted, and dash-dotted lines correspond,
  respectively to Z=0.06 and 13~Gyr, Z=$10^{-4}$ and 1~Gyr, and
  Z=0.019 and 100~Myr. The higher density of {\it random} spectra
  between 7\,000 and 8\,000~K is a consequence of the temperature
  overlap between the two different {\it reference} grids, for hot 
  and cool stars (see Sect.~\ref{Grids}), from which the random
  spectra have been interpolated. }
\label{FigGridRandom}
\end{figure}

For the determination of the code performances, we selected a subsample
of the previously described 20\,000 random spectra, based on the following criteria:
\begin{itemize}
\item  In order to correctly populate the Hertzsprung-Russell diagram
  (in terms of stellar parameters combinations, and not of stellar
  lifetimes) as expected for a galaxy like the Milky Way with
  different stellar populations (assuming a standard initial mass function
  over a long period of star formation), we first retrieved PARSEC
  v1.1 isochrones from \citet{Bressan12}.  The selected isochrones
  correspond to ages of 1 and 13~Gyr and metal content $Z = 10^{-4}$
  and 0.06 (i.e. \meta = -2.2 and +0.7~dex).
\item We then selected among the 20\,000 interpolated spectra
  described in Sect.~\ref{GridRandom}, those having a (\T , \g )
  combination located between these two isochrones.
\item Finally, for any selected (\T , \g ) combination, we chose all
  the available interpolated spectra (whatever their \meta \ and
  \alfaFe \ values are).  These criteria led to the selection of
  9\,067 random spectra.
\item In addition, and to consider the few hot giant stars that could
  be present in the RVS surveyed volume, we also retrieved an
  isochrone that is representative of younger stars (100~Myr) with solar
  metallicity ($Z = 0.19$). We then selected all the giants having 
  (i) \T \ varying between 9\,000 and 11\,000~K and, (ii) a surface gravity comprised
  between 3.5~\gunits \ and the isochrone gravity values. This procedure
  added 898 stars to the total selected sample.
\end{itemize}

{The final high and low resolution {\it random} grids (as called hereafter)
are composed of  9\,965 noised random spectra  at each of the six selected
\GRVS \ or S/N-values. } Their location in the Hertzsprung-Russell
diagram is shown in Fig.~\ref{FigGridRandom}. Their distribution in
metallicity and \alfa -enrichment is perfectly flat over the ranges
[-2.5, +0.5] and [-0.4,+0.8], respectively, because of their random
nature.

\subsection{Influence of the post-launch abandoned RVS spectra rebin
on \gsp \ performances}
\label{Res}

\begin{figure*}[t!]
\includegraphics[height=18cm, width=18.cm]{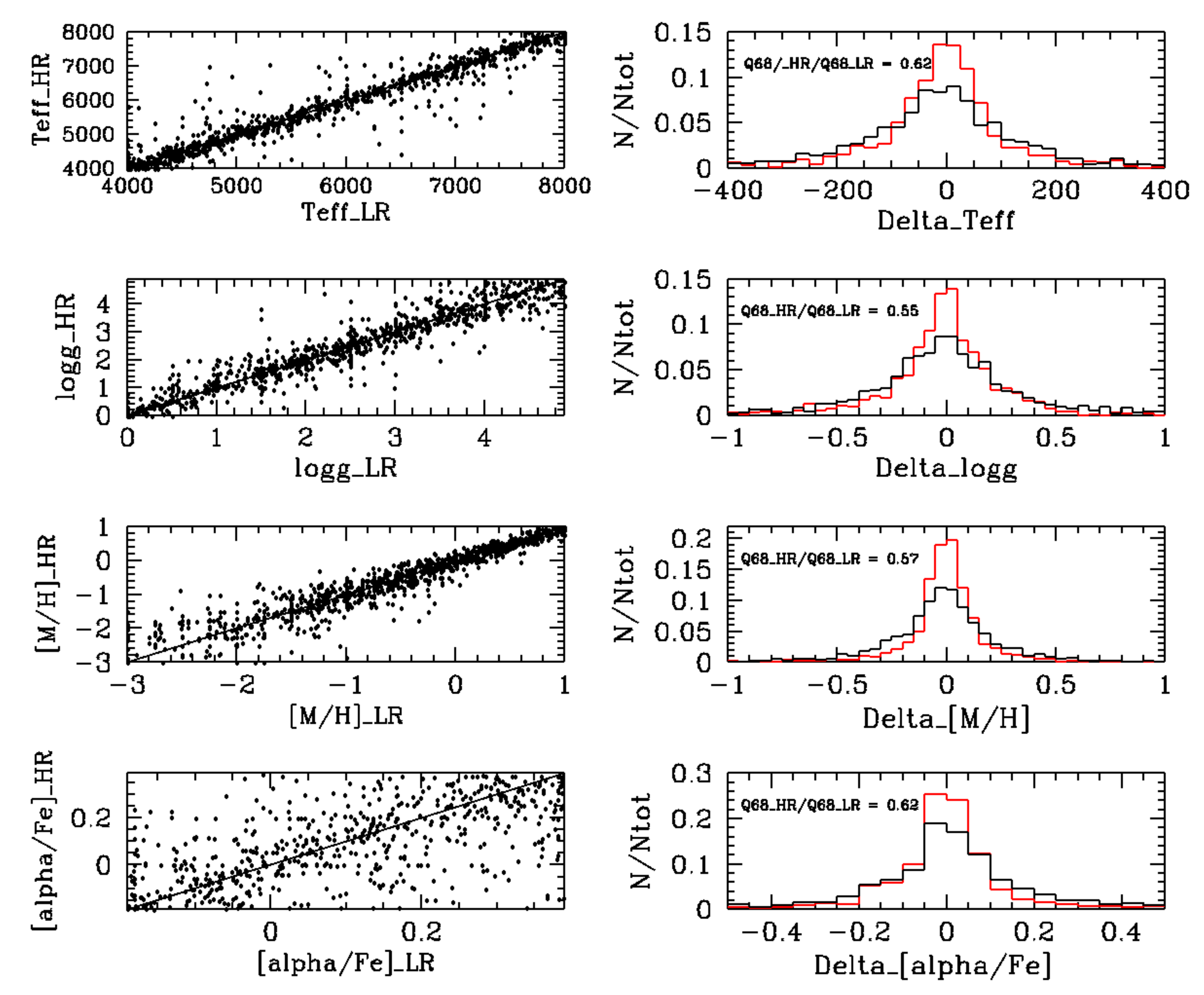}
\caption{Left panels: MATISSE$_G$ results for cool random
spectra of S/N$\sim$20 at the nominal high resolution (HR) compared to those 
for the lower resolution (LR) spectra of the same stars (and same S/N value).
Right panels: normalized distribution of residuals for the high-resolution spectra
(red curve) and for the low-resolution ones (black curve). The ratio of the 68\%
quantile values of both distributions is given for each stellar parameter.}
\label{LRHR}
\end{figure*}

As explained in the Introduction, following the actual RVS performances revealed by the commissioning phase,
it has been decided that every RVS spectra will be provided in the nominal high-resolution
mode to minimize the background contamination.  
Before launch, a binning by a factor three for stars fainter than \GRVS$<$10 was planned \citep[e.g.][]{BailerJones13}, 
decreasing the effective resolution of their spectra to around 7500. 
The parametrization algorithms tested for \gsp \ have therefore faced the problem of correctly
evaluating this recent effective resolution change on their performances. 

In this work, we have decided to
analyse and to report the influence of this post-launch revision, while providing up-to-date performance
expectations in agreement with the actual nominal RVS configuration. The resolution change issue
has been tackled through the following steps:

\begin{itemize}

\item One of the codes, MATISSE$_G$ (already integrated in the Apsis \gaia DPAC pipeline) 
 was run both on all the noised random spectra with the nominal
 resolution of R~$\sim$11\,200, and on their corresponding rebinned spectra
 with effective resolution R~$\sim$7\,500. The four left panels of Fig.~\ref{LRHR} show 
 the MATISSE$_G$ results for cool random spectra of S/N$\sim$20 at the nominal high resolution 
 (HR) compared to those for the rebinned lower resolution (LR) spectra of the same stars and same
 S/N value.
 No particular trends are found between both solutions, justifying the possibility of correcting
 the parametrization performances for this rebinning change in the input data.

\item The influence of the resolution in the parametrization
precision was quantified by estimating the variation of the 68\% quantile of the error 
(residuals) distributions (see also Sect.~\ref{MAR}) due to the resolution change only. The four right panels of Fig.~\ref{LRHR} 
present normalized distribution of residuals for the S/N~$\sim$20 high-resolution  spectra results (red curve)
and for the low-resolution ones (black curve) of cool random stars. The ratio of the 68\% quantile values of both 
distributions is given for each stellar parameter. Similar ratios have been estimated for all the S/N values 
considered in this paper (c.f. Tab.\ref{Tab_LRHR}). In general, we can see that at constant S/N, the gain in the parameters precision 
when passing from the low resolution rebinned spectra to the nominal high resolution ones is of about one third.

\item The results of the FERRE and ANN models which had been trained on rebinned low-resolution spectra for 
S/N 125, 40, 20 and 10 (fainter stars, as expected before launch), could now be rescaled to what we would 
expect if they had been trained on high-resolution spectra. The performances of those codes for
 the fainter stars, quantified through the 68\% quantile of the residuals distribution, have been
 corrected by multiplying by the high-to-low resolution ratios derived in the previous step for each
 stellar parameter and each S/N value.

\end{itemize}

In summary, thanks to the above described synthetic spectra samples that consider the most
up-to-date S/N-magnitude relation, and properly dealing with the actual RVS
configuration and resolution, the parametrization tests presented here can be used
to confidently estimate the future \gsp \ performances.

\begin{table}[t!]
\caption{Ratio of the 68\% quantile values of the error distributions for nominal high resolution data and
 low effective resolution data (in the sense Q68$_{\rm HR}$ divided by Q68$_{\rm LR}$), as estimated with MATISSE$_G$. The results for different parameters and
 S/N values are presented.}
\label{Tab_LRHR}
\centering 
\begin{tabular}{c cccc}
& & & &\\
\hline
& & & &\\
S/N     & 125    & 40  & 20  & 10   \\
\GRVS   &  10.3 & 11.8 & 12.6 & 13.4 \\
& & & &\\
\hline
\T      & 0.62 & 0.66 & 0.62 & 0.62 \\
\g    &  0.44 & 0.48 & 0.55 & 0.51 \\
\meta   & 0.62 & 0.62 & 0.57 & 0.68 \\
\alfaFe  & 0.72 & 0.73 & 0.68 & 0.83 \\
& & & &\\
\hline
\end{tabular}
\end{table}


\section{Performance comparison of the different parametrization codes}
\label{Perf}

We present in this section the performances of the tested methods for
the sample of noised random FGK-type and BA-type stars defined in
Sect.~\ref{RandomSamples}.  In the following, the reported and
discussed performances are those obtained by the FERRE and ANN codes
together with those of MATISSE locally improved by GAUGUIN (noted
MATISSE$_G$, hereafter).  In particular, the erosion of those
codes performances as the information contained in the spectra
decreases (ex. for increasing noise, lack of spectral signatures,...)
is analysed here, in order to understand the behaviour of each
method and its best applicability domain.

\subsection{Distribution of residuals}

\begin{figure*}[ht!]
\includegraphics[height=15.cm,width=18.cm]{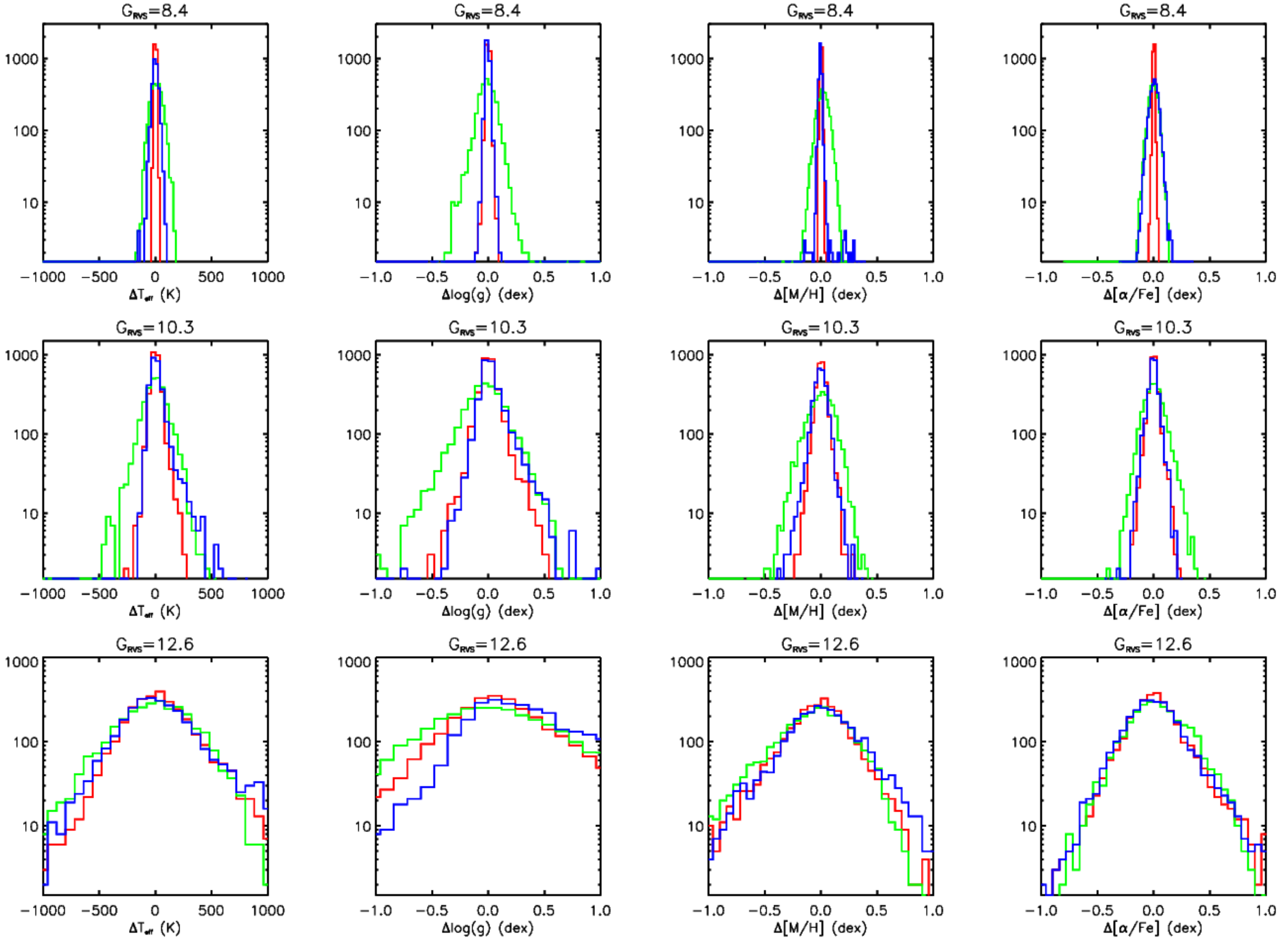}
\caption{Distributions of the residuals in the recovered atmospheric
  parameters ($\Delta \theta = \theta _{rec} - \theta _{real}$) for a
  subsample of cool random spectra with \GRVS = 8.4, 10.3 and, 12.6 (i.e. S/N values of 350, 125 and 20}, from
  top to bottom, respectively) and defined by 4\,000~$<$ \T $<$ and
  8\,000~K. 
  The different colours refer to the different tested methods: FERRE, ANN
  and, MATISSE$_G$ in red, green and blue, respectively.
\label{PerfDistrib1}
\end{figure*}

\begin{figure*}[ht]
\sidecaption
\includegraphics[height=15.cm,width=13.cm]{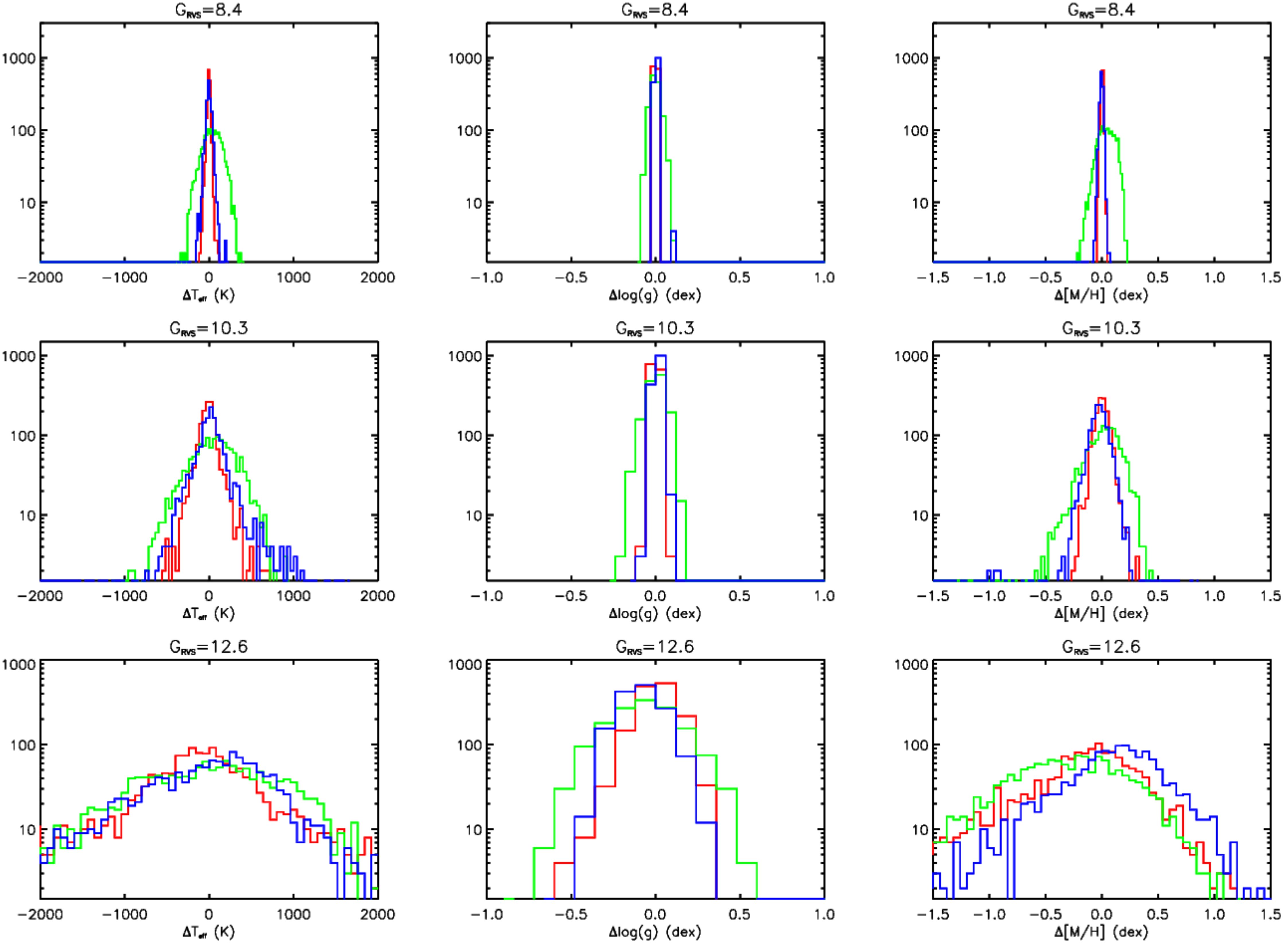}
\caption{Same as Fig.~\ref{PerfDistrib1} but 
for a subsample of hot random spectra defined
by \T $>$ 8\,000~K. }
\label{PerfDistribHot}
\end{figure*}

To evaluate each code performances, we first computed the differences
between the recovered (i.e. {\it estimated}) and real ({\it input})
atmospheric parameters, $\Delta \theta = \theta _{rec} - \theta
_{real}$ with $\theta$ referring to \T , \g , \meta \ and \alfaFe.  The
$\Delta \theta$ will be referred as the {\it residuals}, hereafter.

Fig.~\ref{PerfDistrib1} and Fig.~\ref{PerfDistribHot} show the
distribution of these $\Delta \theta$ residuals obtained with the three
tested methods, for cool and hot stars random spectra, respectively.
Both figures illustrate the results obtained at different \GRVS
\ magnitudes  (\GRVS = 8.4, 10.3 and 12.6, that correspond to S/N
values of  350, 125 and 20). 

It can be seen that the residual distributions are always very
peaked and depart from a perfect Gaussian-like distribution only at
the faintest magnitudes (it has to be taken into account the
fact that both figures are in logarithmic scale). Almost no outliers (spectra that are
parametrized with an error well outside the main distribution) are
seen.  Moreover, these distributions are not biased at all except for
the faintest hot star spectra where small biases appear only for some
methods.

\begin{figure*}[ht]
\includegraphics[height=11.cm,width=18.cm]{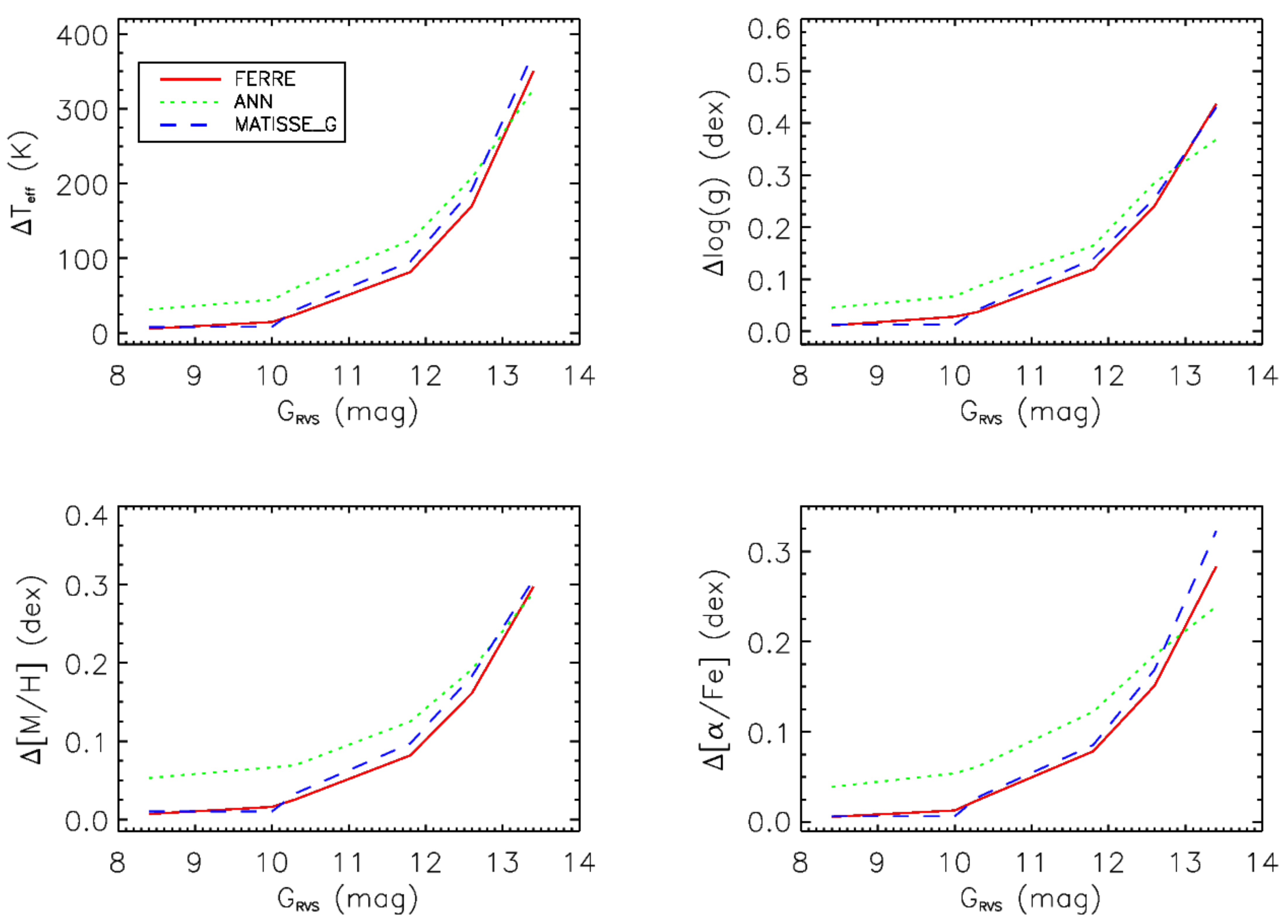}
\caption{Variation of the code performances (quantified by the 68\% quantile) as a function of 
the magnitude, for the subsample of random cool stars defined by 
4\,000~$<$ \T $<$ and   8\,000~K and \meta $\geq$ -1.0~dex (2\,951 spectra in total).}
\label{PerfClass1}
\end{figure*}

Moreover, we can notice that all the methods, at a given magnitude, recover
the four atmospheric parameters with a rather similar
quality. The performances are particularly excellent for the
best-quality spectra (\GRVS = 8.4, this is also true as long as \GRVS
$<$~10, see Fig.\ref{PerfClass1} for instance). The residual
distributions get wider as the noise increases, although the large
majority of the spectra (see, for instance, the discussion on the
Q68$_\theta$ below) are always recovered with an acceptable accuracy,
even at \GRVS = 12.6.  As expected, the parametrisation in \T \ and \meta
\ of the hottest faintest stars is of poorer quality since these spectra
lack of sensitive spectroscopic signatures (see Fig.~\ref{Spec1} where
few atomic lines are seen, except the \ion{Ca}{II} triplet, when \T
$\ga$~8\,000~K).  However, the appearance of the broad Paschen lines
still allows a very good estimate of the stellar surface gravity, even
for very low quality spectra.  In contrast, the surface gravity
of the late-type stars is always the most difficult parameter to recover as
already shown in several previous studies.

\subsection{Quantification of method performances}
\label{MAR}

To quantify each method performances, the 68\% quantile of
the $\Delta \theta$ distributions (Q68$_\theta$, hereafter) was adopted.
This quantile could be viewed as the 1-$\sigma$ error of the parameter
recovery in case of a perfect Gaussian distribution of $\Delta
\theta$. This is however not always the case, in particular, for
low S/N ratios.  Another statistical estimate of the performances is
the systematic error (or bias) that corresponds to the mean of the
differences between the recovered and real parameters ($<\Delta
\theta>$). In our case, the biases are almost always very small
compared to the Q68$_\theta$ quantiles, for every method (see previous
subsection). They can thus be neglected for our purpose and, as a
consequence, they will not be discussed hereafter.

We first point out that we favoured, in the following, these Q68
quantiles over other possible statistical indicators (such as the root
mean square, rms, or the Mean Absolute Residual, MAR) since we believe
that the Q68 are more representative of the real performances (in terms
of the bulk of the tested data) of the
methods, particularly at low S/N ratios where the parametrization is
more difficult to perform. As an illustration of this, we indeed
compared (see Tab.\ref{Tab_MAR}) the MAR$_\theta$ and rms$_\theta$
with the Q68$_\theta$ for all the random FGK-spectral type stars
(about 6\,000 spectra) and three \GRVS \ magnitudes considered in the
present article.  In this table, the reported numbers are obtained
with the FERRE code but our conclusions are independent of the
adopted method.  Although the MAR$_\theta$ and Q68$_\theta$ are very
similar for the best quality spectra (\GRVS $<$ 10.3), it can be seen
that the MAR$_\theta$ are always smaller than the Q68$_\theta$ for
every atmospheric parameter and that this departure increases for
decreasing S/N.  For instance, the MAR$_\theta$ gets close to the
60\% quantiles when  \GRVS \ = 13.4 (S/N$=$10).
This results from the fact that, for fainter spectra, 
the $\Delta \theta$ distributions start to depart
from a pure Gaussian distribution (see 
Fig.~\ref{PerfDistrib1} \& ~\ref{PerfDistribHot}). 
In contrast, the rms$_\theta$ tend to
correspond to higher quantiles than the Q68$_\theta$.
As it is well known, this is caused by the high sensitivity of
the rms to outliers. However, for the
faintest magnitudes, these two statistical indicators become
closer eachothers and the reported errors become in agreement. 


\begin{table*}[t!]
\sidecaption
\caption{Comparison between the statistical performance indicators
  Q68$_\theta$, MAR$_\theta$ and rms$_\theta$ for the FERRE code. We report in the parenthesis of the
  last six columns the quantiles corresponding to the MAR$_\theta$ and the rms$_\theta$.}
\label{Tab_MAR}
\centering 
\begin{tabular}{c ccc c ccc c ccc}
& & & & & & & & & & &\\
\hline
& & & & & & & & & & &\\
         & \multicolumn{3}{c}{Q68$_\theta$} & & \multicolumn{3}{c}{MAR$_\theta$} & & \multicolumn{3}{c}{rms$_\theta$} \\
& & & & & & & & & & &\\
\cline{2-4} \cline{6-8} \cline{10-12}
& & & & & & & & & & &\\
S/N        & 125    & 40    & 10    &  & 125 & 40 & 10 & & 125 & 40 & 10\\
\GRVS   &  10.3 & 11.8 & 13.4 & & 10.3 & 11.8 & 13.4 & & 10.3 & 11.8 & 13.4\\   
& & & & & & & & & & &\\
\T \ (K)     & 32.1 &103.3 &381.8& & 33.4 (Q69) &101.4 (Q68) &327.0 (Q62) & & 56.1 (Q84)&153.7 (Q80)&436.0 (Q73)\\
\g \ (dex)     & 0.05 & 0.15 & 0.49& & 0.05 (Q68) & 0.15 (Q67) & 0.40 (Q61) & & 0.08 (Q82)& 0.22 (Q79)& 0.57 (Q71)\\
\meta \ (dex)  & 0.05 & 0.14 & 0.36& & 0.05 (Q68) & 0.13 (Q65) & 0.29 (Q60) & & 0.08 (Q82)& 0.18 (Q76)& 0.38 (Q71)\\
\alfaFe \ (dex) & 0.04 & 0.14 & 0.35& & 0.04 (Q68) & 0.12 (Q66) & 0.27 (Q58) & & 0.09 (Q83)& 0.18 (Q78)& 0.35 (Q68)\\
& & & & & & & & & & &\\
\hline
\end{tabular}
\end{table*}

\begin{figure}
\includegraphics[height=5.3cm,width=8.4cm]{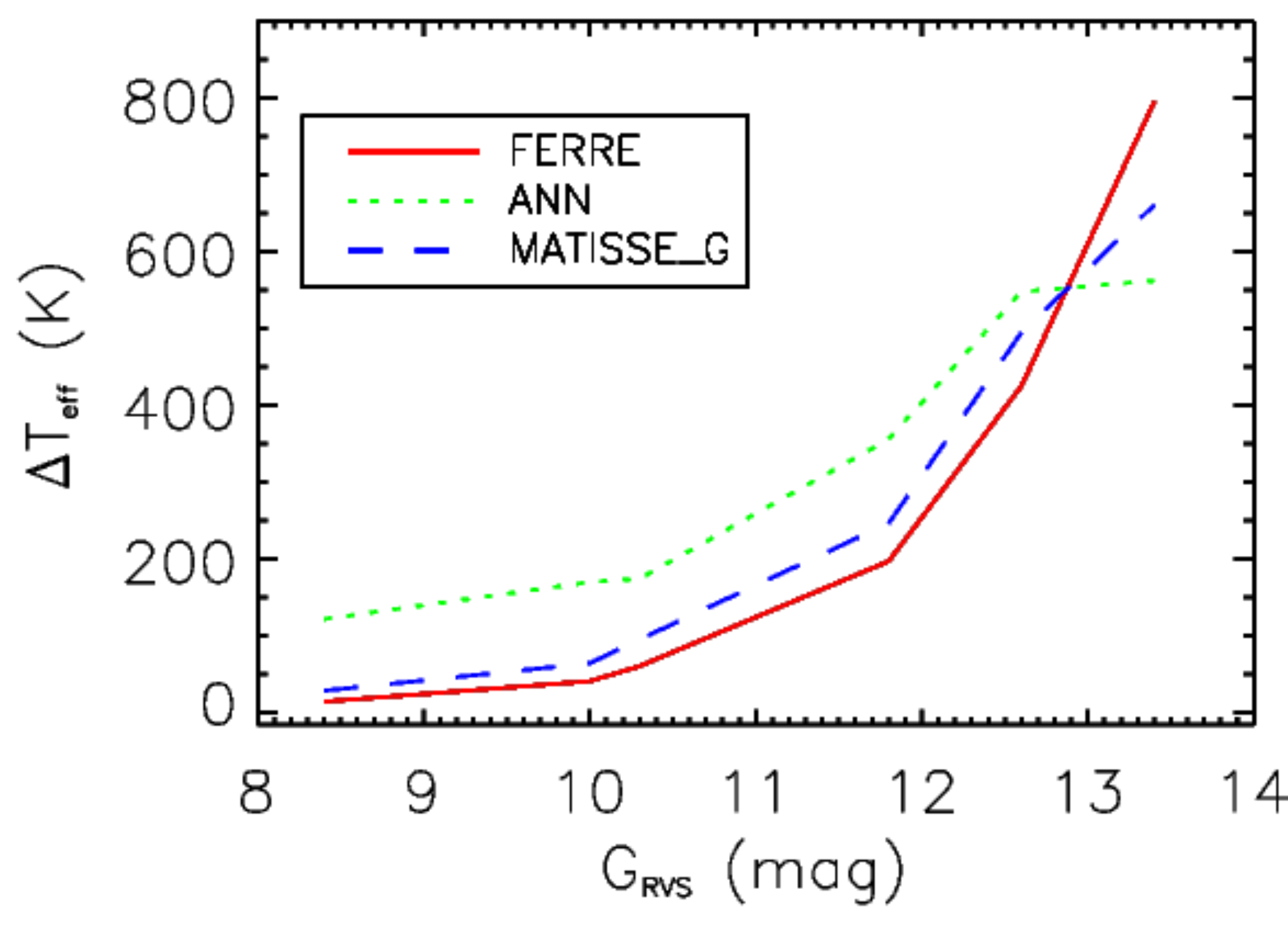}
\includegraphics[height=5.3cm,width=8.4cm]{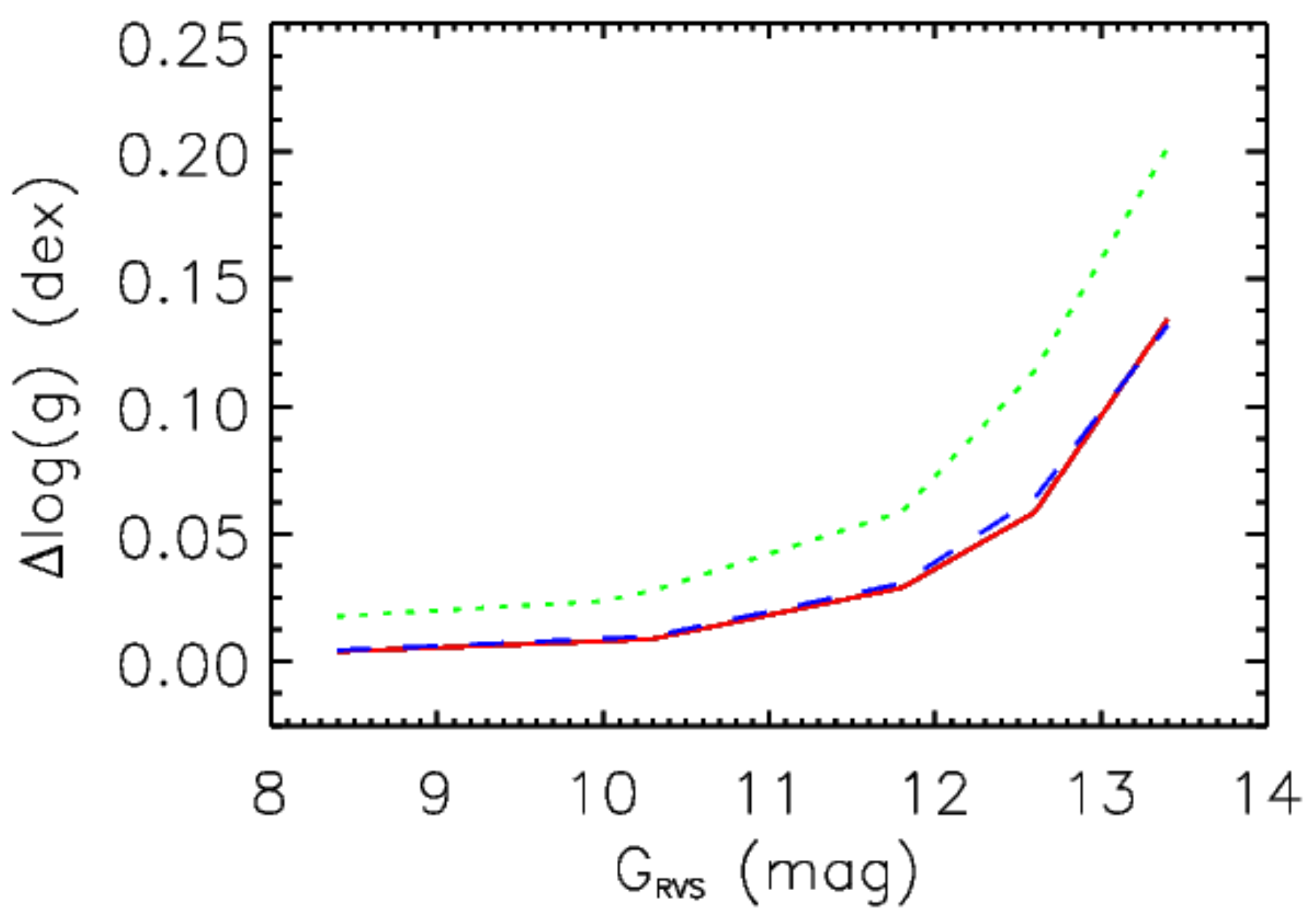}
\includegraphics[height=5.3cm,width=8.4cm]{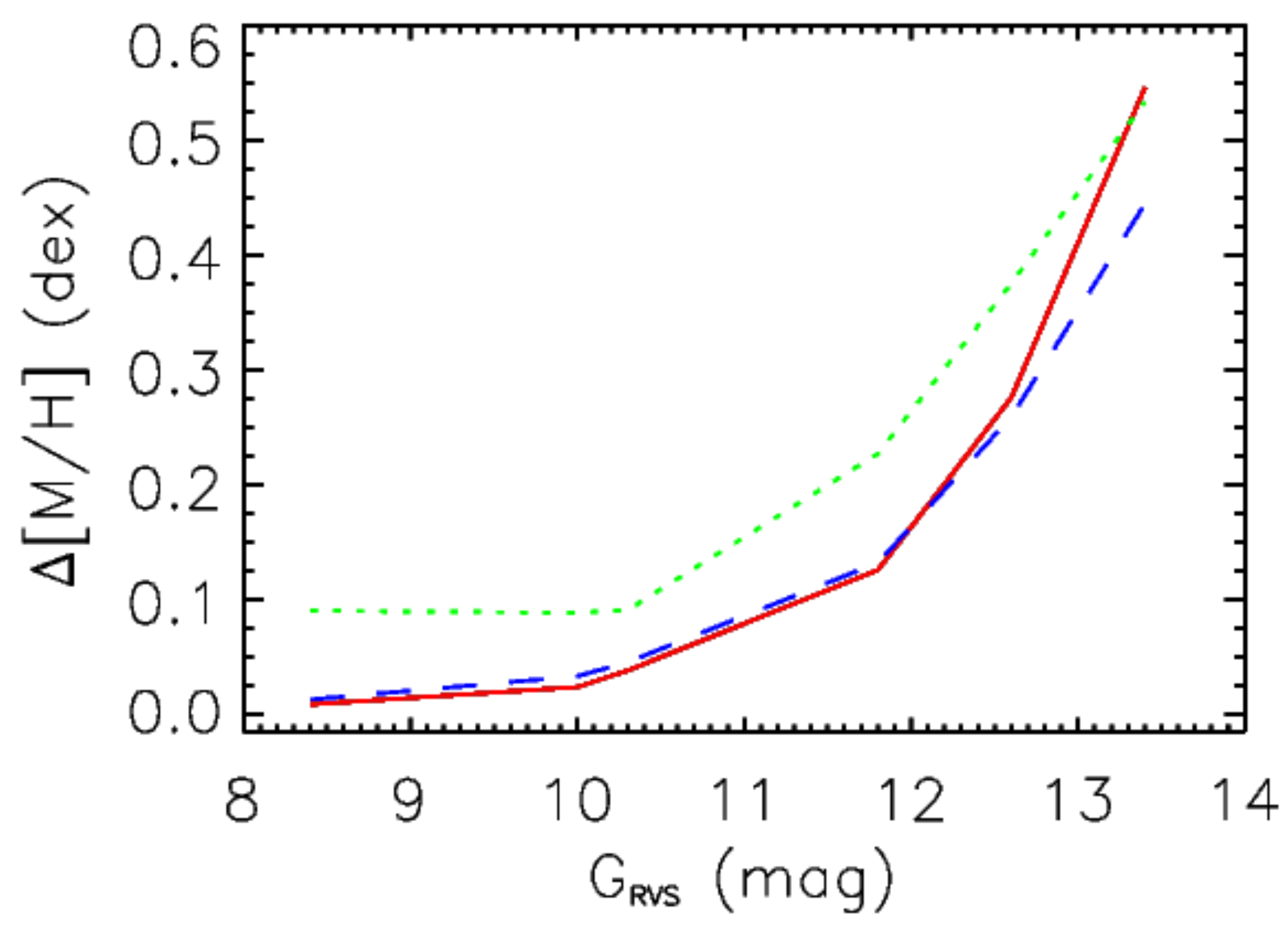}
\caption{Same as Fig.~\ref{PerfClass1} but for the subsample of hot stars, defined by \T $>$ 8\,000~K 
and \meta $\geq$ -1.0~dex (1\,457 spectra in total).}
\label{PerfHotClass1}
\end{figure}

\subsubsection{General cool and hot random samples: noise and
effective temperature effects}
As a first step of the comparison, it is important to understand how
the parametrization codes react to i) S/N degradation and ii) 
the general {\it palette} of spectral types (and therefore stellar
effective temperatures) that the \gsp \ module will have to deal with.
To this purpose, Fig.~\ref{PerfClass1} \& \ref{PerfHotClass1}
illustrate the degradation of each code performances with
increasing noise for the late-type and early-type stars samples, respectively.  

\begin{figure*}[ht]
\includegraphics[height=11.cm,width=18.cm]{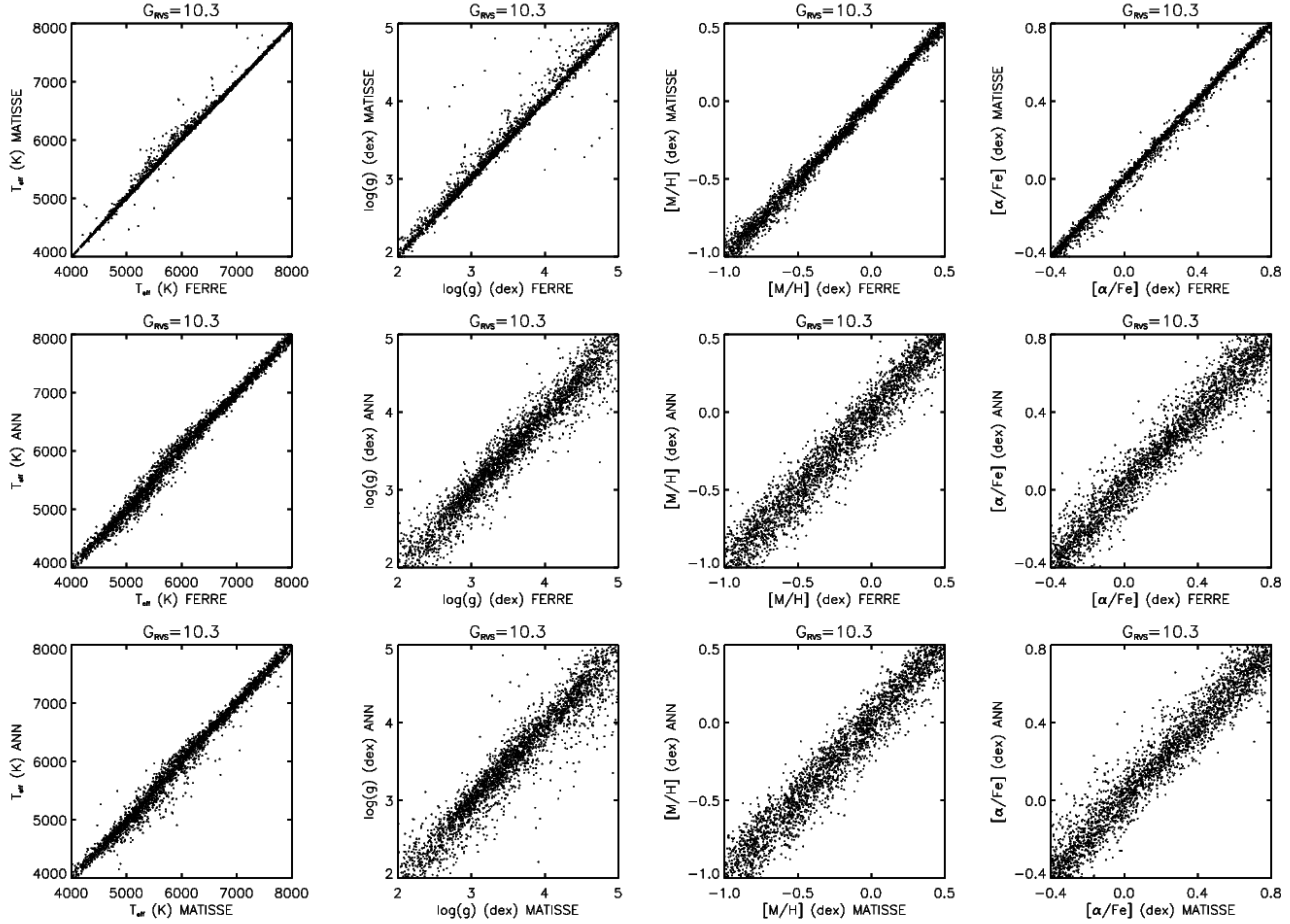}
\caption{Comparison of results  method-to-method, at SNR$\sim125$ (\GRVS$\sim$10.3), for the subsample of random cool stars
defined in Fig.~\ref{PerfDistrib1}.}
\label{PlotScatter1}
\end{figure*}

\begin{figure*}[ht]
\includegraphics[height=11.cm,width=18.cm]{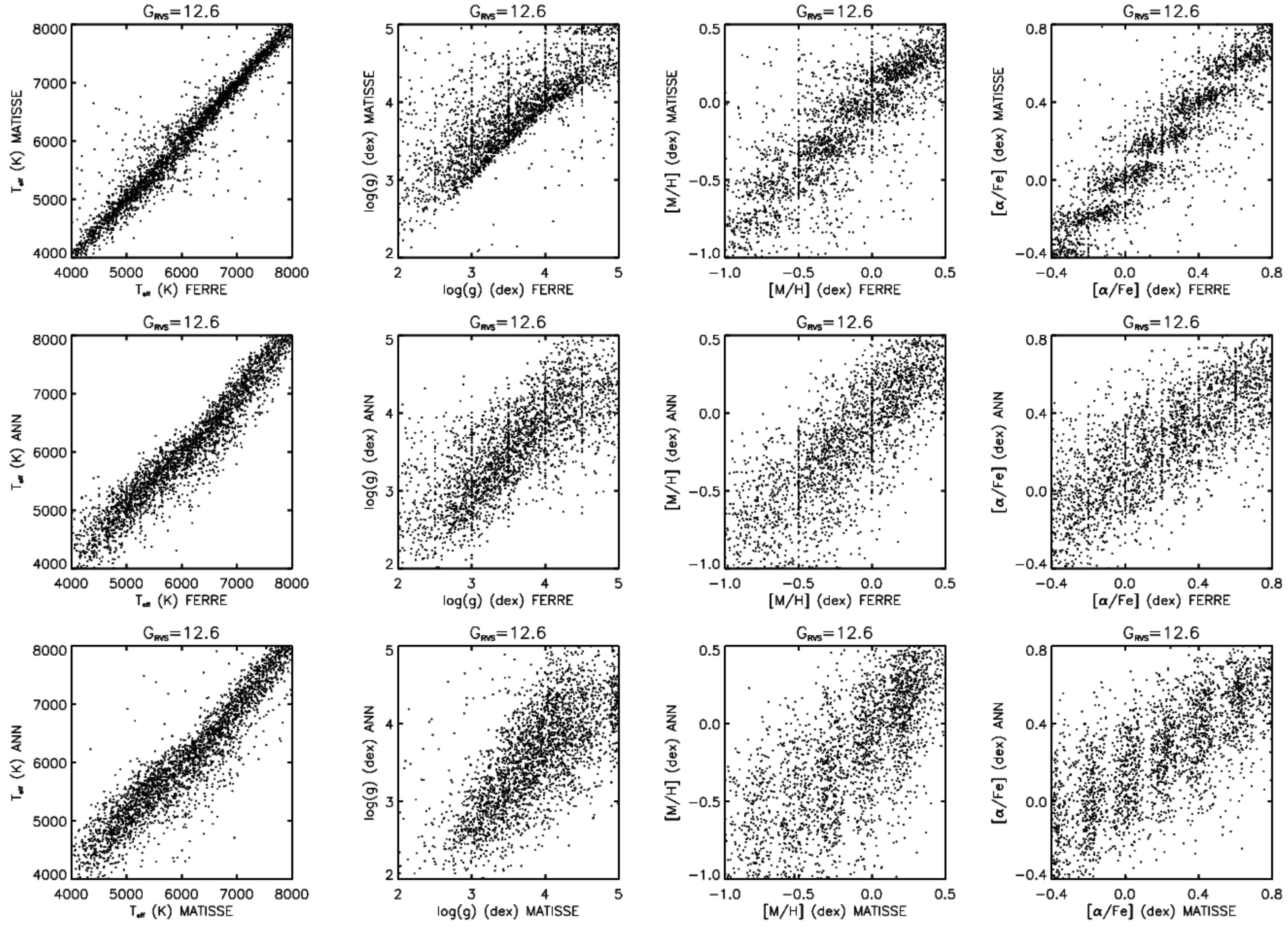}
\caption{Same as Fig.~\ref{PlotScatter1} but at SNR$\sim20$ (\GRVS$\sim$12.6).}
\label{PlotScatter2}
\end{figure*}

In addition, Fig.~\ref{PlotScatter1} and  \ref{PlotScatter2} show the scatter plots of the method-to-method results comparisons 
at SNR$\sim125$ (\GRVS$\sim$10.3) and SNR$\sim20$ (\GRVS$\sim$12.6) respectively. No important trends are found between the
results of the different methods, confirming the consistency, with a higher or lower agreement, between the three types of results.

 From the above mentioned plots, several conclusions can be derived:

\begin{itemize}
\item Most of the stars are well parametrized in \T , \g \ and \meta
  \ (together with \alfaFe \ for the cool stars) by the three methods,
  working completely independently. This reinforces the idea that our
  estimates are robust.
\item In the good to intermediate quality regime  (for \GRVS $\lesssim 12.5$) and
  for cool and hot stars, two completely independent methods (FERRE
  and MATISSE$_G$) produce very compatible results with no
  significative differences between both codes.
\item In the low quality regime (for \GRVS $\gtrsim 12.5$) and cool stars,
  the three methods (FERRE, ANN and MATISSE$_G$) give similar results 
  (see also Fig.~\ref{PerfDistrib1} bottom panel),
  although the ANN method seems to perform slightly better for very
  low S/N spectra  (\GRVS $\sim 13.5$). 
\item For \GRVS $\gtrsim 12.5$  and hot stars, MATISSE$_G$ solutions seem
  slightly more robust.
\end{itemize}

\subsubsection{Gravity and metallicity effects}
\begin{figure*}[ht!]
\includegraphics[height=11.cm,width=18.cm]{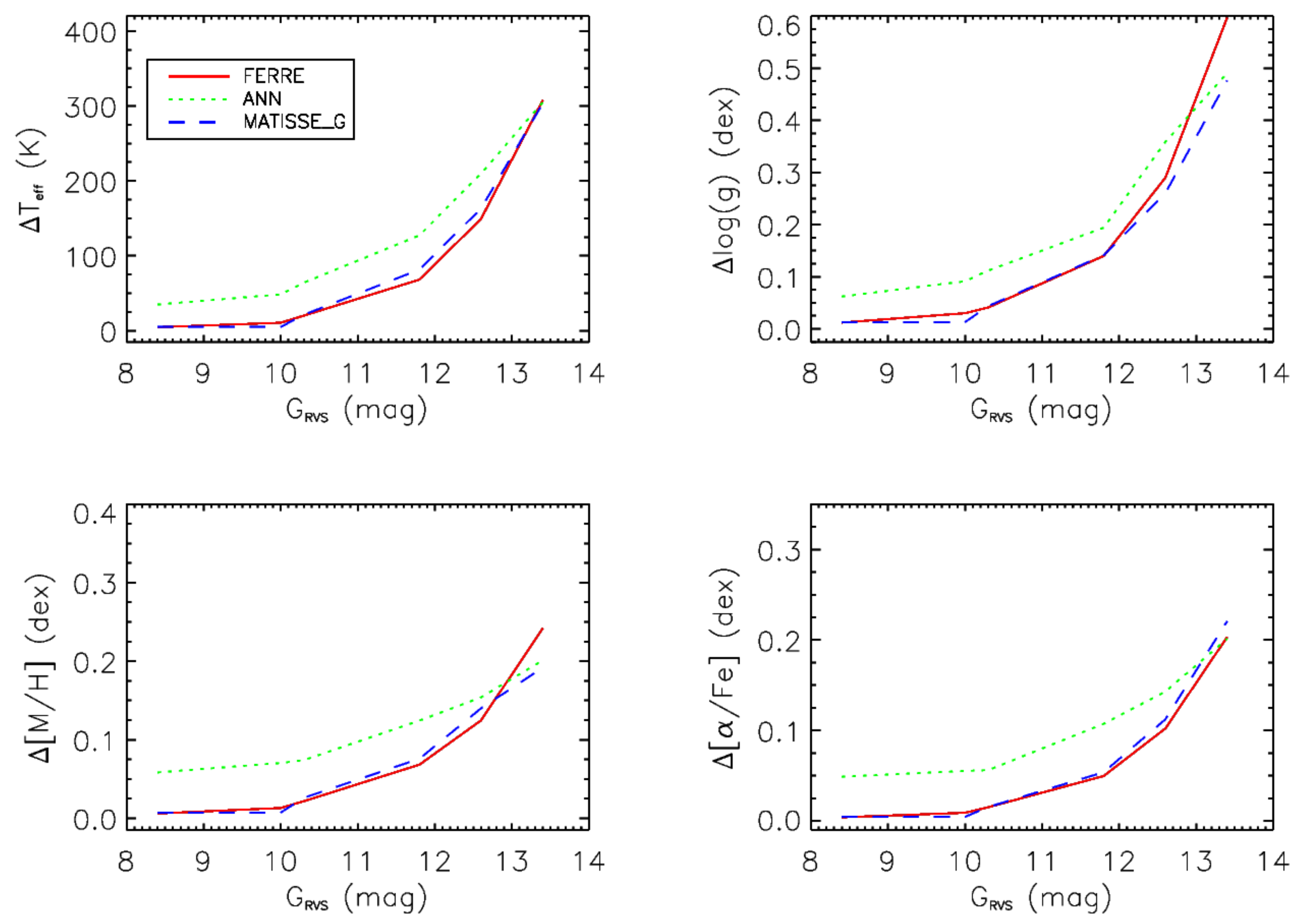}
\caption{Same as Fig.~\ref{PerfClass1} but for a subsample of random cool giant stars defined by \T $<$~6\,000~K,
\g ~$<$~3.5~\gunits \ and \meta ~$\geq$~-0.5~dex (557 stars in total)}
\label{PerfClass4}
\end{figure*}

\begin{figure*}[ht!]
\includegraphics[height=11.cm,width=18.cm]{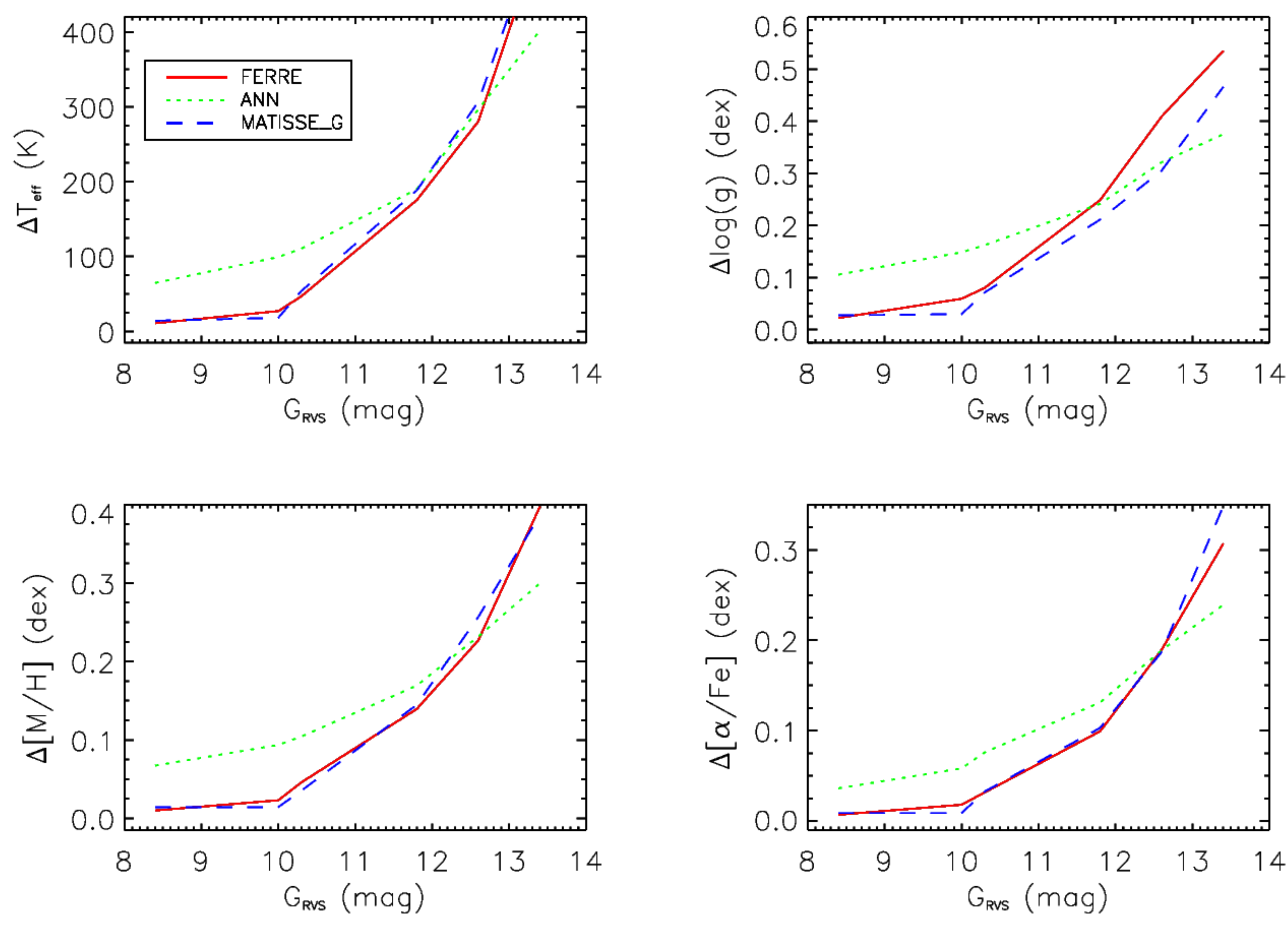}
\caption{Same as Fig.~\ref{PerfClass1} but for a subsample of random cool dwarf stars defined by \T $<$~6\,000~K,
\g $\geq$~3.5~\gunits \ and -1.25$\leq$\meta $<$-0.5~dex (376 stars in total)}
\label{PerfClass6}
\end{figure*}

To complete the robustness evaluation of the tested codes, and to
understand their optimal applicability domains, we need to analyse
their performances as a function of two additional parameters: stellar
metallicity and surface gravity. To this purpose, we have chosen to
illustrate two particular cases, concentrating on G and K-spectral type stars:
rather metal-rich giants (Fig.~\ref{PerfClass4}) and dwarf stars with
intermediate to low metallicities (Fig.\ref{PerfClass6}).
These two stellar types correspond to extreme cases of the \gsp \
performances 
since (i) late-type giants are more easily parametrized than corresponding dwarfs
and (ii) metal-rich star spectra exhibit much more spectral lines that ease their
parametrization (see Sect.~\ref{FGK}). The results show the following tendencies:

\begin{itemize}
\item In the good to intermediate quality regime  (for \GRVS $\lesssim 12.5$),
  the FERRE and MATISSE$_G$ codes are confirmed as the two
  methods providing the best results, independently of the metallicity
  and the gravity.
\item In the low quality regime (for \GRVS $\gtrsim 12.5$) and for metal-rich
  stars, the three methods have similar performances, with ANN and 
  MATISSE$_G$ codes being slightly better for \g.
\item In the low quality regime (for \GRVS $\gtrsim 12.5$) and for metal poor
  stars, the ANN method is again providing the best results.
\end{itemize}

\subsubsection{Summary of each code performances}
\begin{figure*}[ht!]
\includegraphics[height=5.cm,width=18.cm]{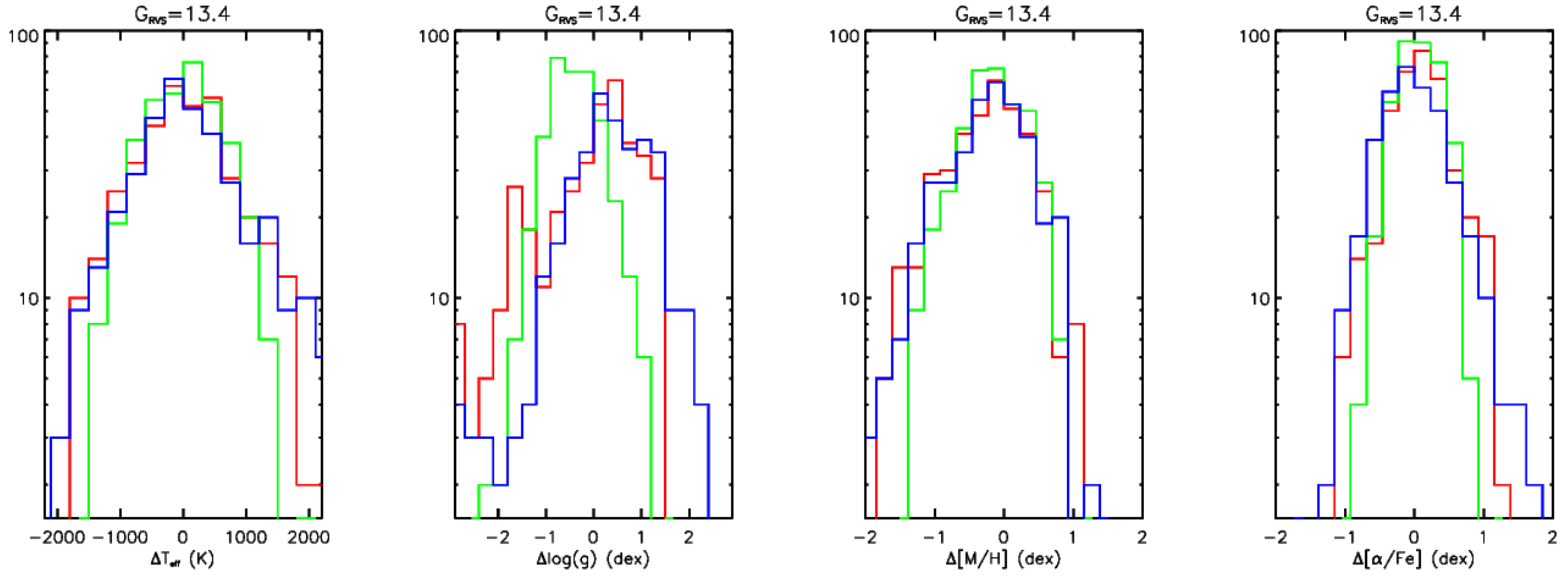}
\caption{Same as Fig.~\ref{PerfDistrib1} but for the subsample of Fig.~\ref{PerfClass6}  and \GRVS = 13.4.}
\label{PerfDistrib2}
\end{figure*}

In summary, from the above performance comparison, we can infer
the following characteristics of each method application to 
the RVS data:
\begin{itemize}
\item The FERRE parametrization is always very satisfactory
  with good results for any parameters of any type of stars.
   However,
  for the faintest spectra, the performances degrade strongly leading
  to rather badly classified cool metal-poor dwarfs.
\item The MATISSE$_G$ method performs rather similarly to FERRE with
  satisfactory parametrisation for every situation.  MATISSE$_G$
  actually produces slightly better results 
 when \GRVS $\la$~10.5 but
  slightly worse for lower S/N ratios. In addition, MATISSE$_G$ can
  sometimes produce the best estimates when the physical parameters
  information is low, although still not completely degraded  at \GRVS
  =~13.5 (as for metal-rich cool giants and early-type stars).  
\item The ANN code always provides the best results for late-type
  stars  when \GRVS $\simeq$~13.5  (see Fig.~\ref{PerfDistrib2}). 
  We stress that such stars will represent the largest
  sample collected by \gaia RVS. 
  We point out that, however, for early-type
  stars, MATISSE$_G$ sometimes performed better.


\item It clearly
  seems that the present version of FERRE and MATISSE$_G$ are very sensitive to the 
  Gaussian-like dominated noise simulated in this work.
  This will be improved in the
  near future with the development of an optimized version of these codes
  when real RVS spectra will be available together with a
  precise knowledge of their noise properties. The adopted filtering for the ANN
  method will also have to be adapted in consequence.
\end{itemize}


\section{Expected parametrization performances for \gaia RVS end-of-mission data}
\label{final}

 From the previous examination of the different parametrisation
codes, we have derived the final (end-of-mission) \gsp \ expected results by
choosing the optimal method for each applicability domain. 
We first point out that the same code solution was adopted for all four (or
three for the early-type stars) atmospheric parameters, to avoid a mix of
physically inconsistent parameters. This selection was performed
through the following main rules, based on the conclusions of
Sect.~\ref{Perf}:

\begin{itemize}
\item For \GRVS $<12.6$ (S/N$> 20$), the average of the FERRE and MATISSE$_G$ solutions has
  been adopted as the final \gsp \ performances since no significant
  differences appear between both methods.
\item  For \GRVS $\geq 12.6$ (S/N$\geq= 20$)  and FGK stars, the average of the FERRE,
  MATISSE$_G$ and ANN is chosen for
  metal-rich and intermediate-metallicity stars, while the ANN method is favoured for late-type metal-poor
  spectra.
\item   \GRVS $\geq 12.6$ (S/N$\geq 20$) and hot stars, MATISSE$_G$ solutions have been
  selected.
\end{itemize}
In any case, the \gsp \ pipeline will also provide the individual results of the different
codes in order to avoid any possible discontinuities between the different parameter and/or S/N regimes.
Such discontinuities could be accidentally
produced by these adopted rules. We however point out that this should be 
avoided owing to an accurate validation phase based on
the analysis of benchmark stars (see an example of such a procedure within GES in
Recio-Blanco et al., 2015, in preparation).

First of all, in order to describe the code performances for
specific types of stars (in terms of spectral type, luminosity and
metallicity), we have defined different stellar classes characterized
by the following atmospheric parameters ranges:

\begin{enumerate}
\item Metallicity ranges:
\begin{itemize}
\item Metal-poor stars: $-2.25 \leq$ \meta \ $< -1.25$~dex, roughly
corresponding to Halo stars.
\item Intermediate-metallicity:  $-1.25 \leq$ \meta \ $< -0.5$~dex, typical of the Galactic thick disk
\item Metal-rich stars:  $-0.5 \leq$ \meta \ $\leq 0.25$~dex, roughly corresponding to the Galactic
thin disc.
\end{itemize}
\item Gravity ranges:
\begin{itemize}
\item Giant stars: 2.5 $\leq$ \g \ $<$ 3.5~\gunits
\item Dwarf stars: 3.5 $\leq$ \g \ $\leq$ 4.5~\gunits
\end{itemize}
\item Effective temperature ranges:
\begin{itemize}
\item B-type stars: 10\,000 $\leq$ \T \ $\leq$ 11\,500~K 
\item A-type stars: 7\,500 $\leq$ \T \ $\leq$ 9\,500~K 
\item F-type stars: 6\,000 $\leq$ \T \ $\leq$ 7\,000~K 
\item G-type stars: 5\,000 $\leq$ \T \ $<$ 6\,000~K
\item K-type stars: 4\,000 $\leq$ \T \ $<$ 5\,000~K
\end{itemize}
\end{enumerate}

This led to the 30 stellar classes (15 for dwarfs and 15 for cool
giants)\footnote{Some of these stellar classes correspond to rather
infrequent real stars, particularly for the hot ones.}.
The end-of-mission \gsp \ parametrization performances for these different classes and \GRVS \ magnitudes
are presented in Tab.~\ref{TabPerfGSPspec}. The following subsections
analyse and discuss the obtained results.

\begin{sidewaystable*}
\caption{Expected end-of-mission performances for \gsp \ (quantified by the 68\% quantile) for the different
stellar classes defined in Sect.~\ref{final}.}
\label{TabPerfGSPspec}
\centering
\begin{tabular}{l  rrrrrr  cccccc  cccccc  cccccc}
\hline \hline
 & &&&&&& &&&&& &&&&& \\
 & \multicolumn{6}{c}{\T \ (K)} & \multicolumn{6}{c}{\g \ (dex)} & \multicolumn{6}{c}{\meta \ (dex)} 
 & \multicolumn{6}{c}{\alfaFe \ (dex)} \\
 & &&&&&& &&&&& &&&&& \\
\hline
 & &&&&&& &&&&& &&&&& \\
S/N        & 350 & 150  & 125 & 40 & 20 & 10 & 350 & 150  & 125 & 40 & 20 & 10 & 350 & 150  & 125 & 40 & 20 & 10 & 350 & 150  & 125 & 40 & 20 & 10\\
\GRVS (mag) & 8.4&10.0&10.3&11.8&12.6&13.4   & 8.4&10.0&10.3&11.8&12.6&13.4 & 8.4&10.0&10.3&11.8&12.6&13.4 & 8.4&10.0&10.3&11.8&12.6&13.4 \\
 & &&&&&& &&&&& &&&&& \\
 \hline
 & &&&&&& &&&&& &&&&& \\
DWARFS & &&&&&& &&&&& &&&&& \\
B metal-rich &   35  &  89   & 138  & 382  & 478  & 744   & 0.01 & 0.01  & 0.01 & 0.02 & 0.04 & 0.11  & 0.01 & 0.03  & 0.05 & 0.13 & 0.32 & 0.51 &&&&& \\
B interm. met. & 36  & 105   & 144  & 420  & 490  & 808   & 0.01 & 0.01  & 0.01 & 0.02 & 0.05 & 0.12  & 0.02 & 0.05  & 0.07 & 0.19 & 0.38 & 0.58 &&&&&\\
B metal-poor &   42  & 108   & 149  & 429  & 499  & 784   & 0.01 & 0.01  & 0.02 & 0.03 & 0.05 & 0.12  & 0.02 & 0.06  & 0.10 & 0.27 & 0.45 & 0.65 &&&&&\\
 & &&&&&& &&&&& &&&&& \\
A metal-rich &   8   &  21   &  31  & 101  & 210  & 323   & 0.01 & 0.01  & 0.01 & 0.03 & 0.07 & 0.12  & 0.01 & 0.01  & 0.02 & 0.07 & 0.14 & 0.18 &&&&& \\
A interm. met. & 8   &  21   &  35  & 104  & 226  & 353   & 0.01 & 0.01  & 0.01 & 0.03 & 0.07 & 0.12  & 0.01 & 0.02  & 0.03 & 0.11 & 0.26 & 0.34 &&&&& \\
A metal-poor &   9   &  24   &  35  & 104  & 232  & 393   & 0.01 & 0.01  & 0.01 & 0.04 & 0.07 & 0.14  & 0.01 & 0.03  & 0.05 & 0.17 & 0.33 & 0.45 &&&&& \\
 & &&&&&& &&&&& &&&&& \\
F metal-rich &   7  &  19    &  34  &  98  & 179  & 336   & 0.01 & 0.03  & 0.04 & 0.10 & 0.15 & 0.26  & 0.01 & 0.02  & 0.02 & 0.08 & 0.12 & 0.20  & 0.01 & 0.01 & 0.02 & 0.06 & 0.14 & 0.26\\
F interm. met. & 10  &  23   &  40  & 134  & 198  & 394   & 0.01 & 0.03  & 0.05 & 0.12 & 0.16 & 0.27  & 0.01 & 0.03  & 0.06 & 0.14 & 0.20 & 0.39  & 0.01 & 0.02 & 0.05 & 0.14 & 0.24 & 0.27 \\
F metal-poor &  13  &  30    &  51  & 134  & 198  & 425   & 0.02 & 0.05  & 0.06 & 0.14 & 0.17 & 0.29  & 0.02 & 0.05  & 0.08 & 0.21 & 0.23 & 0.43  & 0.02 & 0.05 & 0.09 & 0.22 & 0.25 & 0.29\\
 & &&&&&& &&&&& &&&&& \\
G metal-rich &  7  &  19     &  26  &  99  & 166  & 385   & 0.02 & 0.05  & 0.06 & 0.17 & 0.21 & 0.27  & 0.01 & 0.01  & 0.02 & 0.08 & 0.13 & 0.20  & 0.01 & 0.01 & 0.02 & 0.07 & 0.15 & 0.22\\
G interm. met. & 14  &  33   &  62  & 178  & 295  & 460   & 0.03 & 0.08  & 0.12 & 0.18 & 0.25 & 0.28  & 0.01 & 0.03  & 0.06 & 0.16 & 0.25 & 0.32  & 0.01 & 0.02 & 0.04 & 0.10 & 0.20 & 0.27 \\
G metal-poor &  23  &  69    & 105  & 200  & 326  & 487   & 0.05 & 0.13  & 0.15 & 0.22 & 0.28 & 0.32  & 0.02 & 0.06  & 0.11 & 0.22 & 0.31 & 0.38  & 0.01 & 0.04 & 0.06 & 0.14 & 0.18 & 0.25\\
 & &&&&&& &&&&& &&&&& \\
K metal-rich &  4  &   7     &  22  &  44  & 143  & 255   & 0.01 & 0.04  & 0.06 & 0.12 & 0.22 & 0.28  & 0.01 & 0.01  & 0.02 & 0.06 & 0.11 & 0.20  & 0.01 & 0.01 & 0.02 & 0.04 & 0.15 & 0.19\\
K interm. met. & 8  &  16    &  27  &  74  & 219  & 305   & 0.02 & 0.05  & 0.06 & 0.13 & 0.25 & 0.30  & 0.01 & 0.02  & 0.02 & 0.06 & 0.14 & 0.28  & 0.01 & 0.01 & 0.03 & 0.06 & 0.16 & 0.19 \\
K metal-poor & 19  &  47     &  80  & 183  & 250  & 422   & 0.04 & 0.15  & 0.15 & 0.20 & 0.28 & 0.33  & 0.02 & 0.05  & 0.08 & 0.17 & 0.27 & 0.36  & 0.01 & 0.03 & 0.04 & 0.15 & 0.19 & 0.22 \\
 & &&&&&& &&&&& &&&&& \\
 \hline
 & &&&&&& &&&&& &&&&& \\
GIANTS & &&&&&& &&&&& &&&&& \\
B metal-rich &  11  &  31    &  32  & 107  & 365  & 450   & 0.01 & 0.01  & 0.01 & 0.04 & 0.06 & 0.12  & 0.01 & 0.02  & 0.03 & 0.12 & 0.28 & 0.38 &&&&& \\
B interm. met. &13  &  32    &  45  & 139  & 390  & 473   & 0.01 & 0.01  & 0.01 & 0.04 & 0.06 & 0.12  & 0.01 & 0.03  & 0.06 & 0.19 & 0.40 & 0.45 &&&&& \\
B metal-poor &  16  &  33    &  52  & 194  & 411  & 493   & 0.01 & 0.01  & 0.01 & 0.04 & 0.06 & 0.12  & 0.02 & 0.05  & 0.09 & 0.27 & 0.43 & 0.48 &&&&& \\
 & &&&&&& &&&&& &&&&& \\
A metal-rich &   7  &  27    &  33  &  97  & 209  & 369   & 0.01 & 0.01  & 0.01 & 0.04 & 0.07 & 0.12  & 0.01 & 0.02  & 0.02 & 0.07 & 0.13 & 0.17 &&&&& \\
A interm. met. & 9  &  27    &  34  & 102  & 229  & 382   & 0.01 & 0.01  & 0.01 & 0.04 & 0.07 & 0.14  & 0.01 & 0.02  & 0.04 & 0.12 & 0.22 & 0.30 &&&&& \\
A metal-poor &  11  &  27    &  36  & 110  & 235  & 403   & 0.01 & 0.01  & 0.01 & 0.04 & 0.08 & 0.14  & 0.01 & 0.04  & 0.04 & 0.17 & 0.31 & 0.43 &&&&& \\
 & &&&&&& &&&&& &&&&& \\
F metal-rich &  5  &  10     &  18  &  63  & 134  & 253   & 0.01 & 0.02  & 0.03 & 0.08 & 0.16 & 0.22  & 0.01 & 0.01  & 0.02 & 0.06 & 0.13 & 0.16  & 0.01 & 0.01 & 0.02 & 0.07 & 0.14 & 0.25\\
F interm. met. & 6  &  16    &  22  &  69  & 147  & 265   & 0.01 & 0.03  & 0.03 & 0.11 & 0.17 & 0.22  & 0.01 & 0.03  & 0.04 & 0.11 & 0.17 & 0.22  & 0.01 & 0.02 & 0.04 & 0.12 & 0.22 & 0.27\\
F metal-poor & 6  &  17      &  24  &  71  & 148  & 280   & 0.01 & 0.04  & 0.04 & 0.11 & 0.18 & 0.32  & 0.02 & 0.05  & 0.09 & 0.21 & 0.20 & 0.24  & 0.02 & 0.05 & 0.10 & 0.21 & 0.28 & 0.26\\
 & &&&&&& &&&&& &&&&& \\
G metal-rich & 5  &  12      &  22  &  92  & 177  & 350   & 0.01 & 0.03  & 0.04 & 0.15 & 0.22 & 0.36  & 0.01 & 0.01  & 0.02 & 0.08 & 0.14 & 0.16  & 0.01 & 0.01 & 0.02 & 0.06 & 0.13 & 0.23\\
G interm. met. & 10  &  26   &  47  & 166  & 254  & 373   & 0.02 & 0.06  & 0.10 & 0.21 & 0.30 & 0.44  & 0.01 & 0.03  & 0.05 & 0.16 & 0.23 & 0.25  & 0.01 & 0.02 & 0.04 & 0.12 & 0.19 & 0.24\\
G metal-poor & 15  &  43     &  54  & 170  & 265  & 383   & 0.04 & 0.11  & 0.12 & 0.24 & 0.34 & 0.44  & 0.02 & 0.06  & 0.07 & 0.18 & 0.27 & 0.31  & 0.01 & 0.04 & 0.06 & 0.17 & 0.21 & 0.26\\
 & &&&&&& &&&&& &&&&& \\
K metal-rich &  5  &  11     &  21  &  64  & 147  & 237   & 0.02 & 0.04  & 0.06 & 0.17 & 0.29 & 0.43  & 0.01 & 0.01  & 0.02 & 0.06 & 0.12 & 0.17  & 0.01 & 0.01 & 0.02 & 0.04 & 0.10 & 0.20 \\
K interm. met. &  6  &  18   &  29  &  91  & 211  & 333   & 0.02 & 0.06  & 0.08 & 0.21 & 0.33 & 0.50  & 0.01 & 0.02  & 0.03 & 0.11 & 0.19 & 0.28  & 0.01 & 0.01 & 0.03 & 0.09 & 0.17 & 0.24\\
K metal-poor &  13  &  39    &  65  & 200  & 289  & 444   & 0.04 & 0.10  & 0.13 & 0.28 & 0.34 & 0.52  & 0.02 & 0.05  & 0.07 & 0.19 & 0.25 & 0.31  & 0.01 & 0.04 & 0.05 & 0.14 & 0.19 & 0.26\\
 & &&&&&& &&&&& &&&&& \\
\hline \hline
 & &&&&&& &&&&& &&&&& \\ 
\end{tabular}
\end{sidewaystable*}

\begin{figure*}[ht!]
\sidecaption
\includegraphics[height=9.cm,width=14.cm]{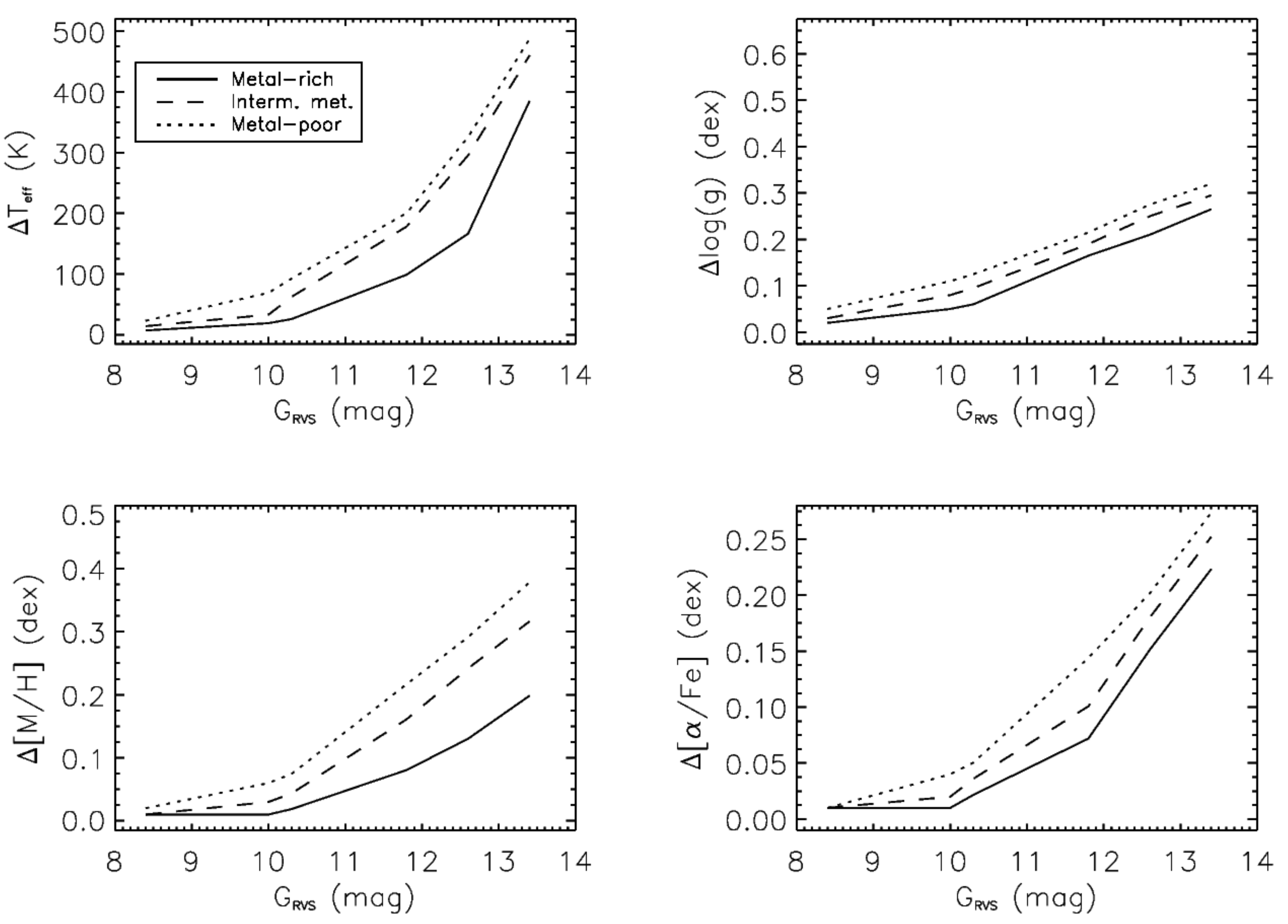}
\caption{Variation of the end-of-mission \gsp \ performances (quantified by the 68\% quantile
of the residuals) as a function of 
increasing magnitudes for the G-dwarf stars defined in Sect.~\ref{final}.}
\label{FigPerfGSPspecDwarf}
\end{figure*}
\begin{figure*}[ht!]
\sidecaption
\includegraphics[height=9.cm,width=14.cm]{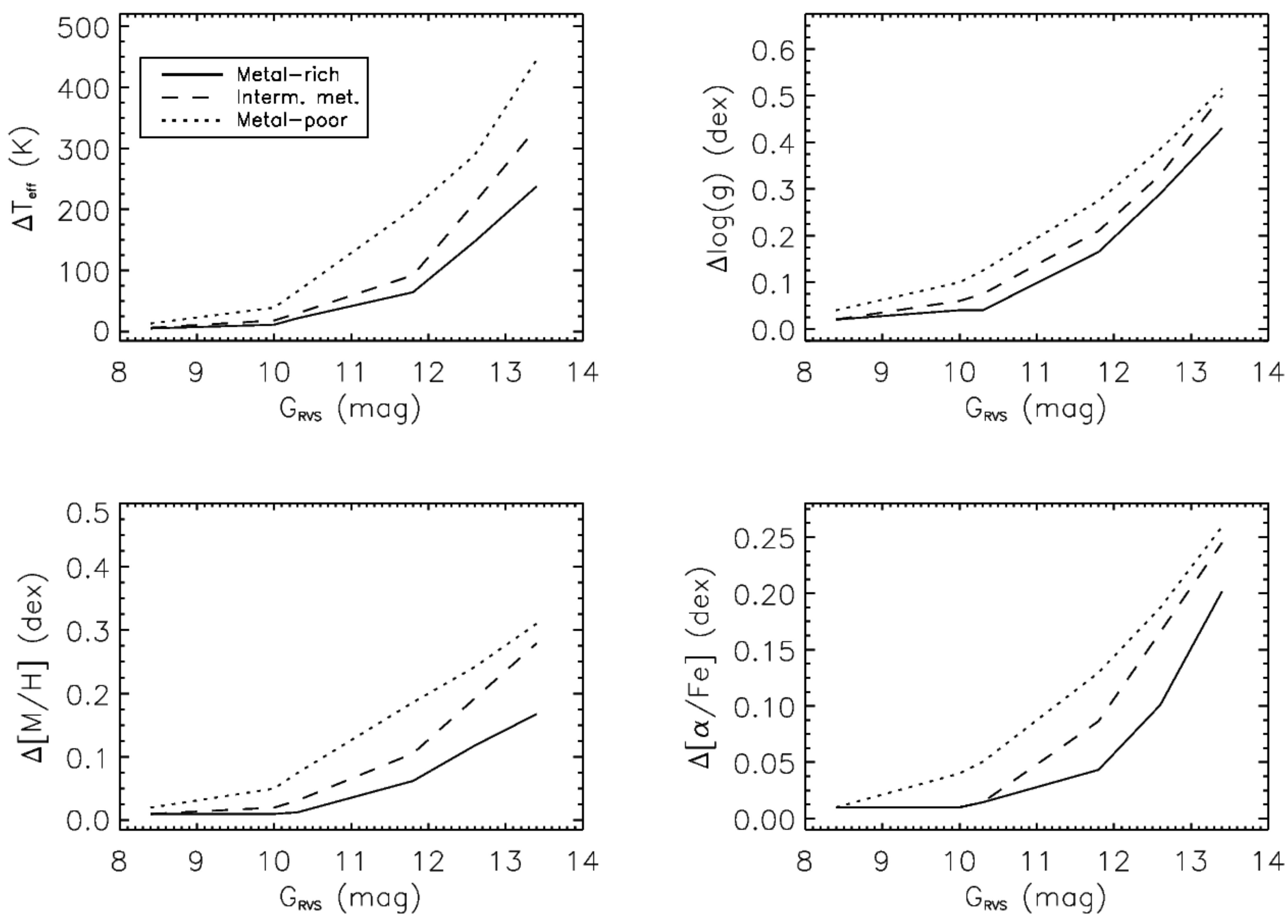}
\caption{Same as Fig.~\ref{FigPerfGSPspecDwarf} but for the K-giant
  stars defined in Sect.~\ref{final}.}
\label{FigPerfGSPspecGiant}
\end{figure*}
\begin{figure*}[ht!]
\sidecaption
\includegraphics[height=8.cm,width=14.cm]{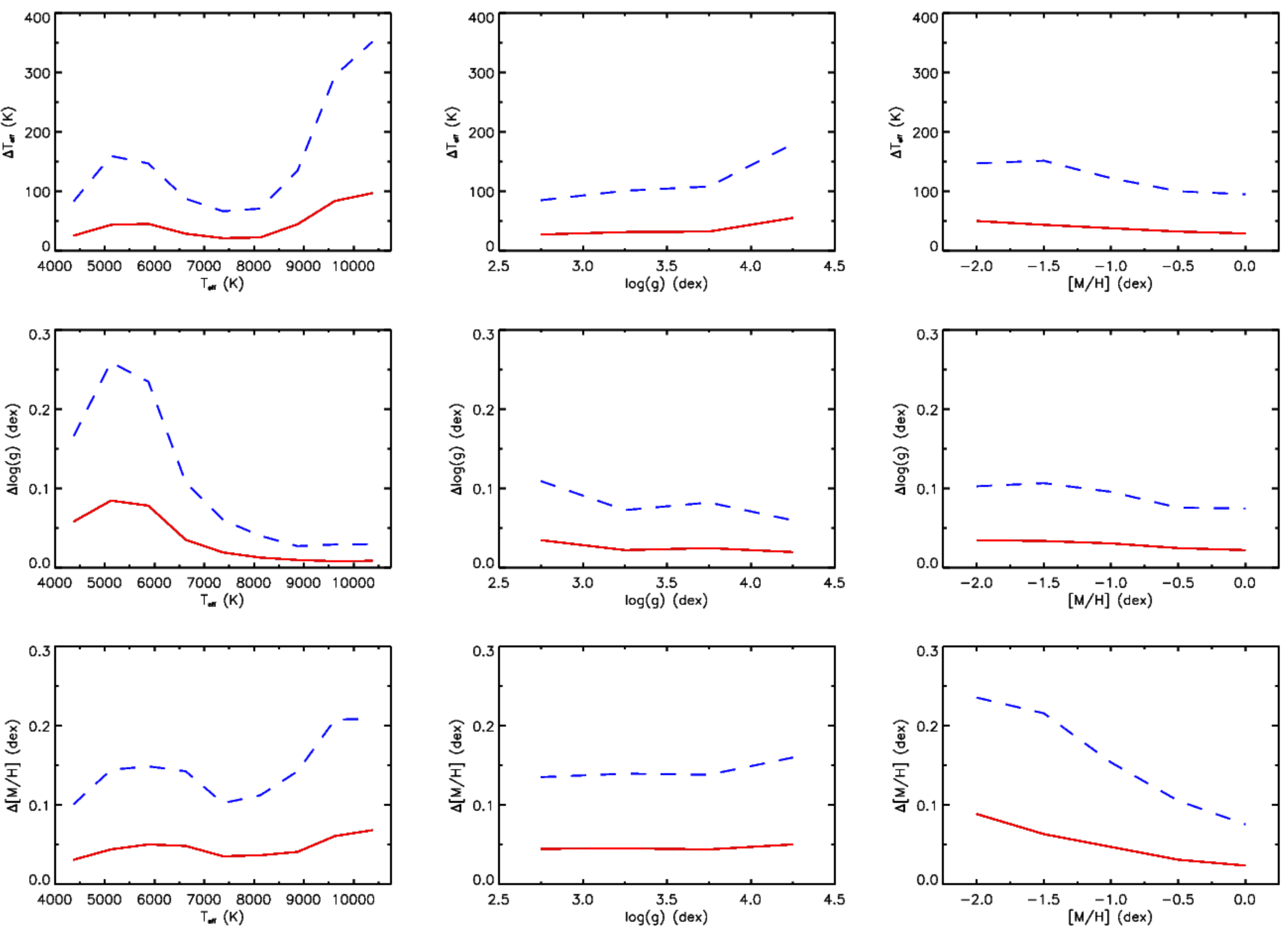}
\caption{Variation of the 68\% quantile of the residuals for \T , \g ,
  \meta \ for early- and late-type stars (end-of-mission \gsp \ performances) as a function of the real
  atmospheric parameters  for \GRVS = 10.3 and 11.8 (S/N $=$ 125 and 40)  in solid red and dashed
  blue line, respectively.  The adopted bins in \T , \g , \meta \ are
  750~K, 0.5~dex \ and 0.5~dex, respectively.}
\label{FigErreur}
\end{figure*}
\begin{figure}[ht!]
\includegraphics[height=4.cm,width=8.7cm]{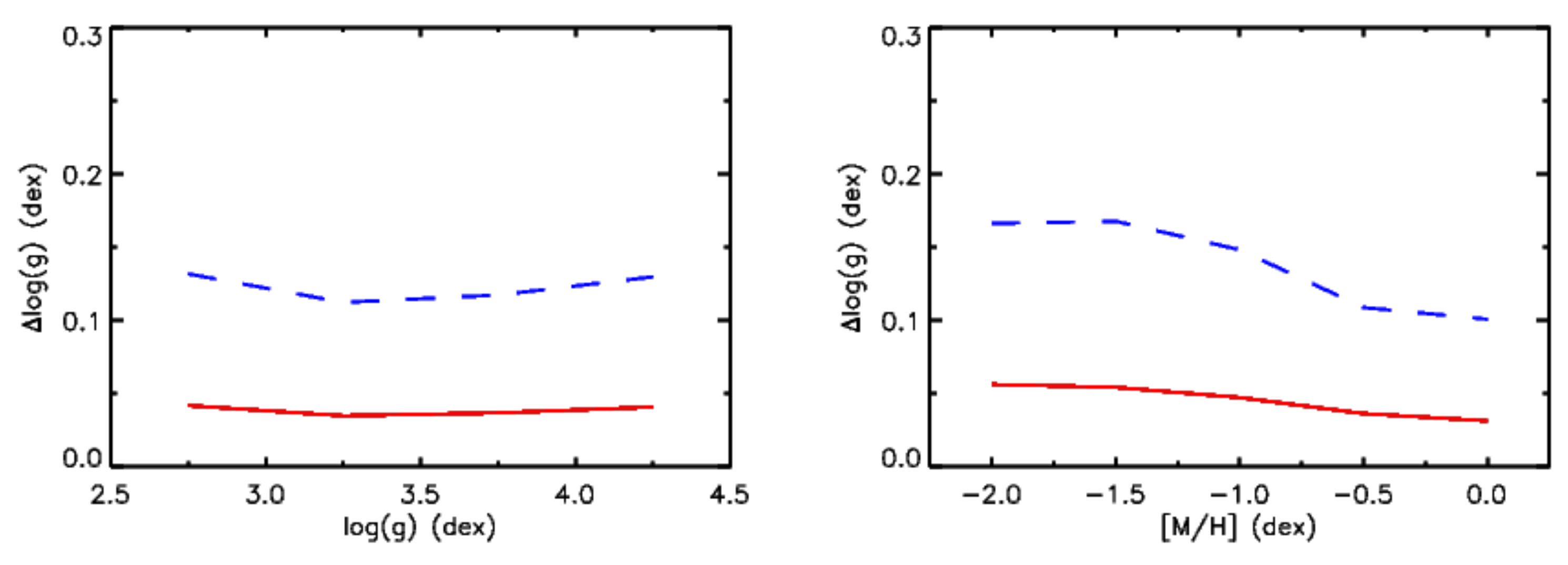}
\includegraphics[height=4.cm,width=8.7cm]{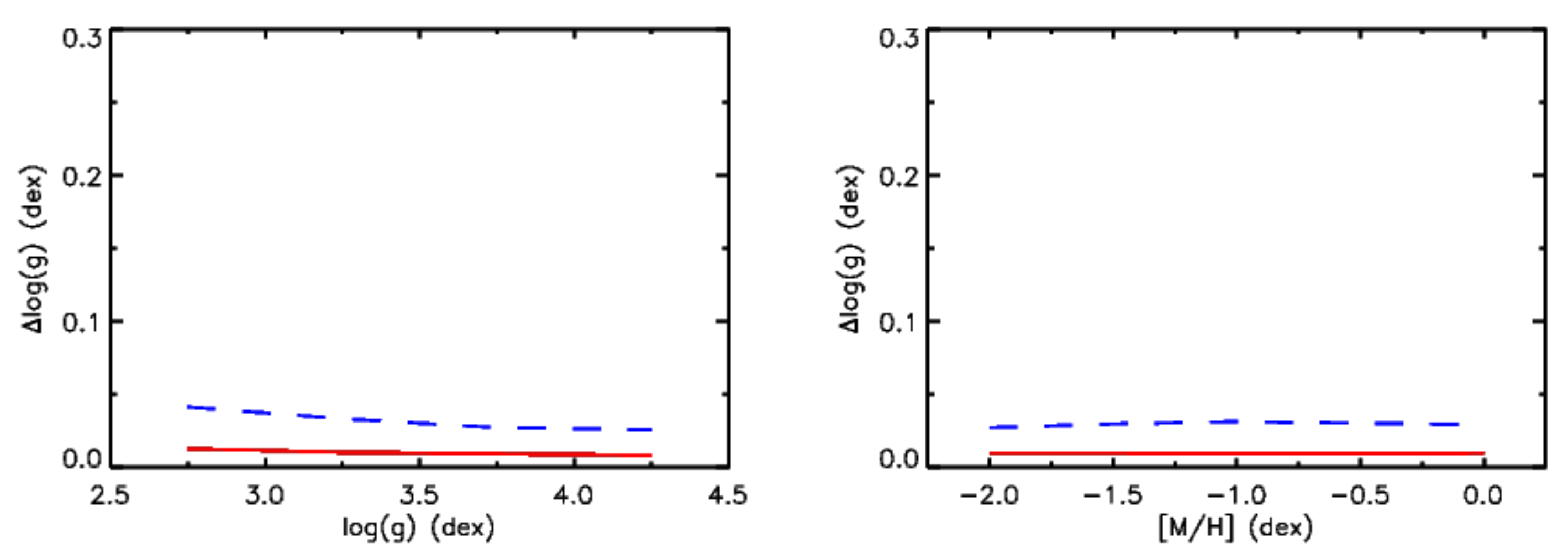}
\caption{Same as Fig.~\ref{FigErreur} but for the residuals in \g ,
  for cool-FGK and hot-BA stars separately (upper and lower panels,
  respectively).}
\label{FigErreurlogg}
\end{figure}

\subsection{Performances for late-type stars}
\label{FGK}
FGK-spectral type stars will represent the majority of the \gaia RVS targets
and therefore, special attention has to be given to the \gsp
\ parametrisation capabilities of their spectra.
Figures~\ref{FigPerfGSPspecDwarf} and \ref{FigPerfGSPspecGiant}
illustrate the expected errors (defined as the 68\% quantile of the
$\Delta \theta$ distributions, c.f. Sect.~\ref{Perf}) for G-type
dwarfs and K-type giants, respectively. The different curves on each panel 
correspond to the three metallicity intervals defined above
and reported in Tab.~\ref{TabPerfGSPspec}.

For stars with \GRVS $\la$12.5, the FGK stars parametrisation is
accurate enough to precisely characterize the stellar properties
(typical errors are smaller than 0.1~dex in \meta \ and \alfaFe) and,
therefore, to conduct Galactic population studies as already performed
from ground-based Galactic archaeology surveys \citep[see][for
instance]{Recio-Blanco14}. Such accuracy in the stellar atmospheric
parameters will allow, in a second step, quite accurate determinations
of individual chemical abundances \citep[see, for
instance,][]{Guiglion14}.  
This is specially true for metal rich and intermediate metallicity stars,
that will be the most abundant ones in the magnitude volume sounded by
the RVS.
For the faintest stars at \GRVS $\geq
12.5-13.0$, where the noise amplitude becomes too strong with respect to the
available stellar spectroscopic signatures, the accuracy in the
parameters estimation degrades (with metallicity errors in the range
0.2 to 0.5~dex, for instance).

On the other hand, the dependency of the parameters accuracy on the
stellar metallicity is illustrated by the clear separation of the
continuous (metal-rich), dashed (metal-intermediate) and dotted
(metal-poor) curves. As expected, metal-rich stars are more easily
parametrized than the cool metal-poor ones whatever the S/N ratios
are. This is evidently caused by the number of spectroscopic
signatures (lines sensitive to the atmospheric parameters and
abundances) available to perform the spectral analysis that
dramatically decreases below \meta $\la$-0.5~dex.  This is also
illustrated in Figures~\ref{FigErreur} and \ref{FigErreurlogg},
showing the dependences of the errors in \T , \g \ and \meta \ on the
three atmospheric parameters. First of all, the three right panels of
Fig.~\ref{FigErreur}, show how the errors in \T \ (right upper panel),
\g \ (right middle panel) \ and \meta \ (right botom panel) depend on
the stellar metallicity.  In each case, two different curves are
shown: the evolution of the 68\% quantile for \GRVS = 10.3 (red
continuous line, S/N$=125$) and for \GRVS =11.8 (blue dashed line, S/N$=40$).  As expected,
the tendencies are more clearly appreciated at \GRVS =11.8 as the
parametrization is more sensitive to loss (or gain) of information at
lower S/N ratios. 

First, it can be appreciated that the errors in the
three parameters increase as the metallicity decreases, with a higher
sensitivity for the metallicity error itself. In addition, the errors
in \T \ increase as the metallicity decreases down to about
\meta=$-1.5$~dex. For lower stellar metallicities, the error in \T
\ remains almost constant and practically independent of \meta. This is
because for those metal-poor stars, the only useful temperature
indicator that remains is the CaII triplet, which is present even at
very low metallicities (see Fig.~\ref{Spec2}). A similar behaviour can be appreciated for the
\g \ errors as a function of \meta \ (right middle panel).  However, in
order to distinguish possible differences between FGK-type and 
early-type stars, Fig.~\ref{FigErreurlogg} shows the evolution of the errors
in \g \ and \meta \ on the same two parameters, but separating FGK and
early-type stars.  The surface gravity error shows in fact a clear
metallicity dependence (right upper panel of Fig.~\ref{FigErreurlogg})
down to \meta=$-1.5$~dex, and no dependence for lower metallicity
stars.

Another important physical parameter influencing the parametrization
performances is the effective temperature. This is illustrated in the
three left panels of Fig.~\ref{FigErreur}. In the range concerning
FGK-type stars (\T \ approximately between 4\,000 and 7\,000~K), the
behaviour of errors in \T , \g \ and \meta \ shows a maximum around
$\sim$5\,500~K. From that point, the errors decrease in both
directions, that is for lower and higher \T. In the first case (for
lower \T \ stars), molecular signatures start to be visible in the
spectra, being more abundant as the temperature decreases. Those
molecular signatures are sentitive to both \T \ and \g, as molecules
formation is favoured for lower temperatures and higher gas pressure
(and therefore \g). In the second case (for stars with \T \ higher
than about 6\,000~K), the appearance of the hydrogen Pachen lines and
their rapid change with \T \ brings a precious gravity indicator that
reduces the errors in \g \ and breaks the \T-\g \ degeneracy
(c.f. Sect.~\ref{hot} and Sect.~\ref{correlations}). As a consequence,
the derivation of \T \ and \meta \ are also improved for those hot stars.

Finally, the gravity influence on the stellar parametrization is
illustrated by the middle panels of Fig.~\ref{FigErreur}. The \T
\ (upper middle panel) and the \meta \ (bottom middle panel) derivation
seem more difficult for dwarf stars than for giants, while the
behaviour seems different for the \g \ estimation (central panel). In
practice, the left panels of Fig.~\ref{FigErreurlogg} showing the
residuals of \g \ as a function of \g \ for FGK-type (upper left
panel) and early-type stars (bottom left one), clarify the situation. For FGK
stars, the gravity determination is in fact also more difficult for
dwarfs than for giants, following the same tendency than the
temperature and metallicity derivations. More generally, in both cases
(FGK dwarfs and giants), the gravity is the more difficult parameter
to estimate. This problem with the surface gravity is mostly caused by
the lack of neutral and ionized lines of the same element in the RVS
spectral domain.  In any case, even for the lowest quality RVS spectra,
the dichotomy between dwarf and giant stars will still be distinguishable.

In summary, the \gaia RVS data of FGK-type stars will allow accurate
studies of the Galactic disc and halo populations. In particular, for
stars with metallicity higher than around $-0.5$~dex, that will be the
majority of the RVS survey, the metallicity and \alfa -enrichment
estimates will be very accurate, with typical errors smaller than
0.1~dex in \alfaFe \ down to \GRVS $\sim$12.5 (a few tens of million
of stars). In addition, K-giants will allow to perform Galactic
studies up to distances of $\sim$5~kpc (for \GRVS=12, and low extinction
regions), or even $\sim$12~kpc (for \GRVS=13.5).

\subsection{Performances for early-type stars}
\label{hot}

\begin{figure}[ht!]
\sidecaption
\includegraphics[height=4.cm,width=7.cm]{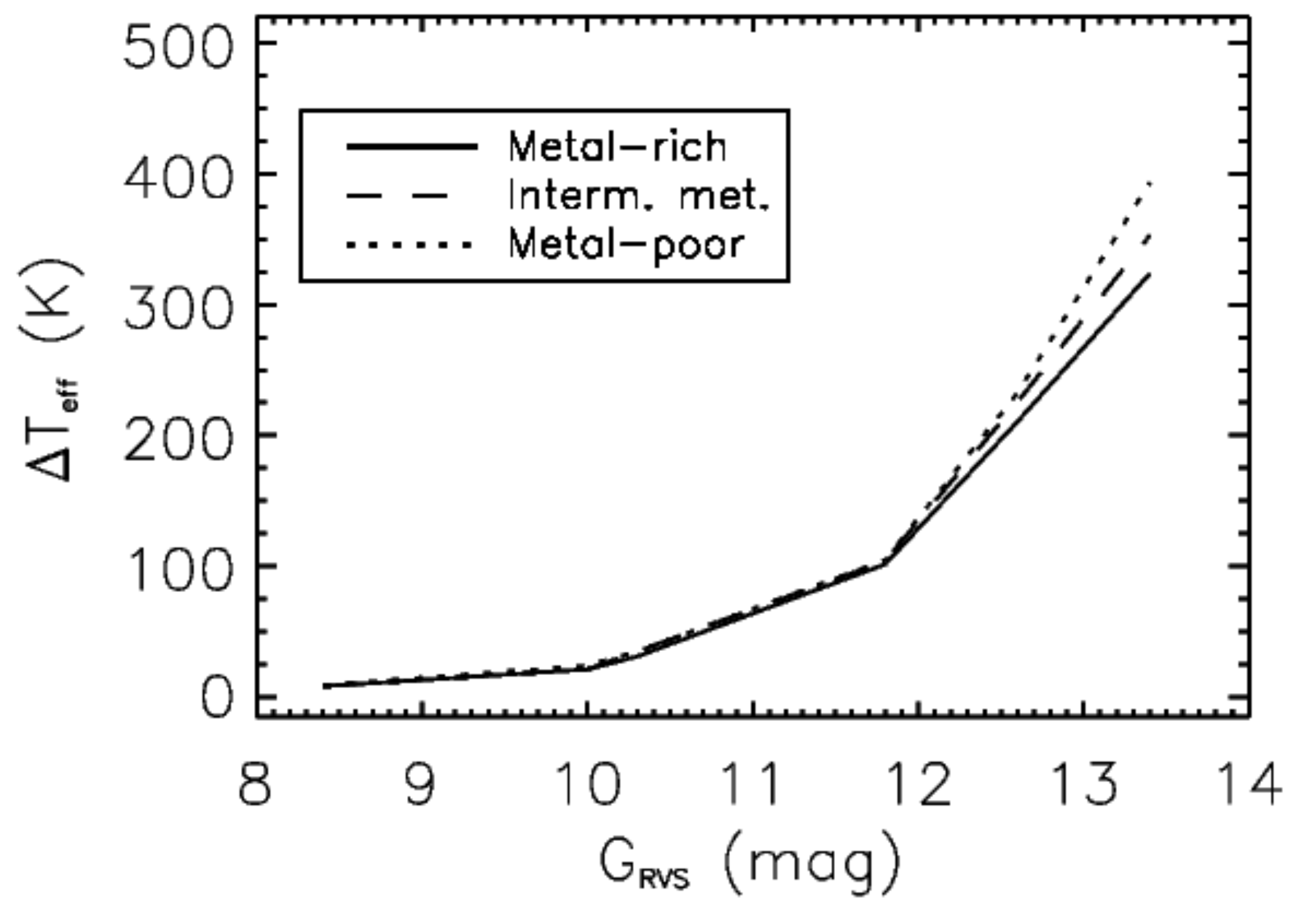}
\includegraphics[height=4.cm,width=7.cm]{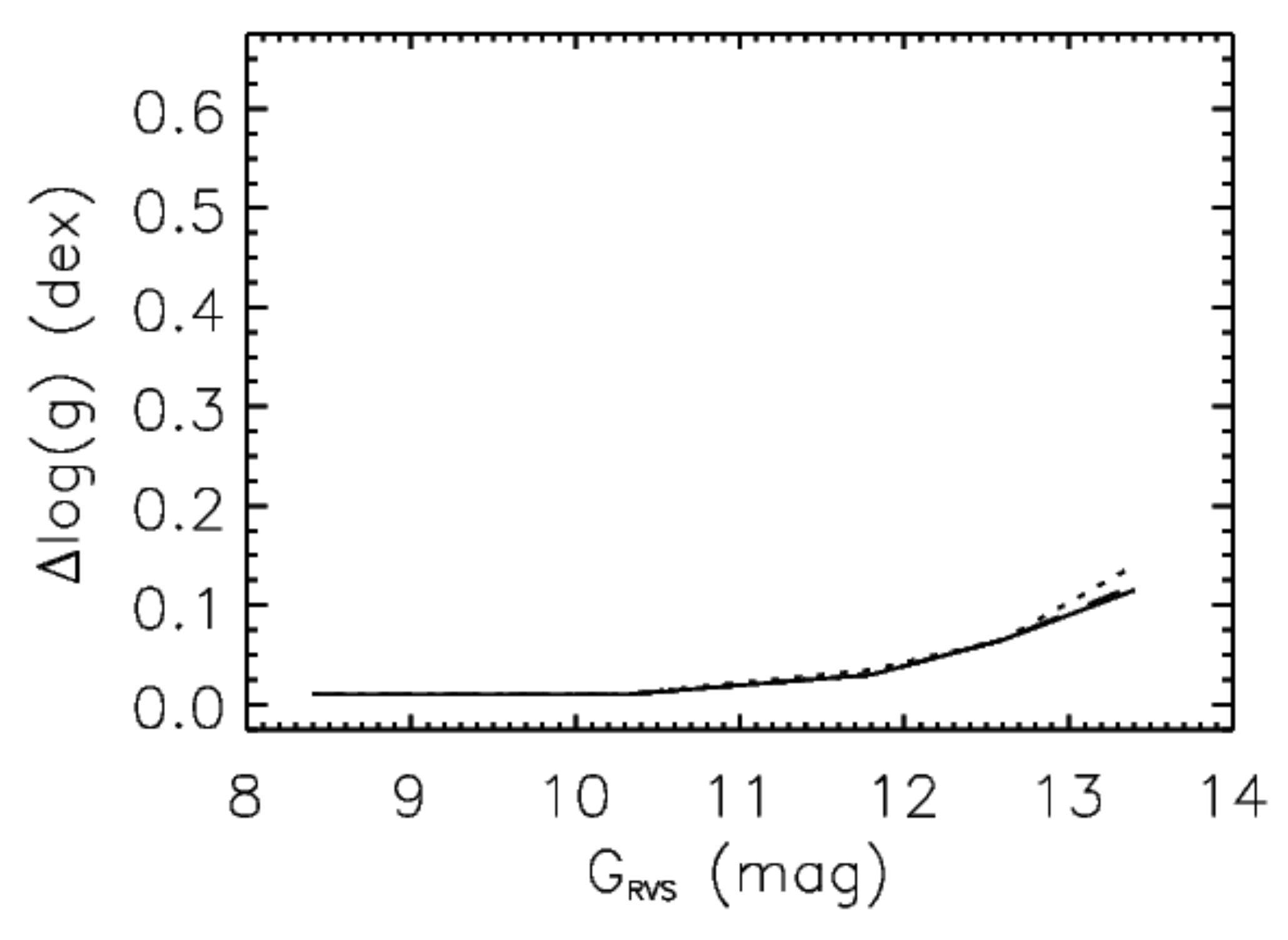}
\includegraphics[height=4.cm,width=7.cm]{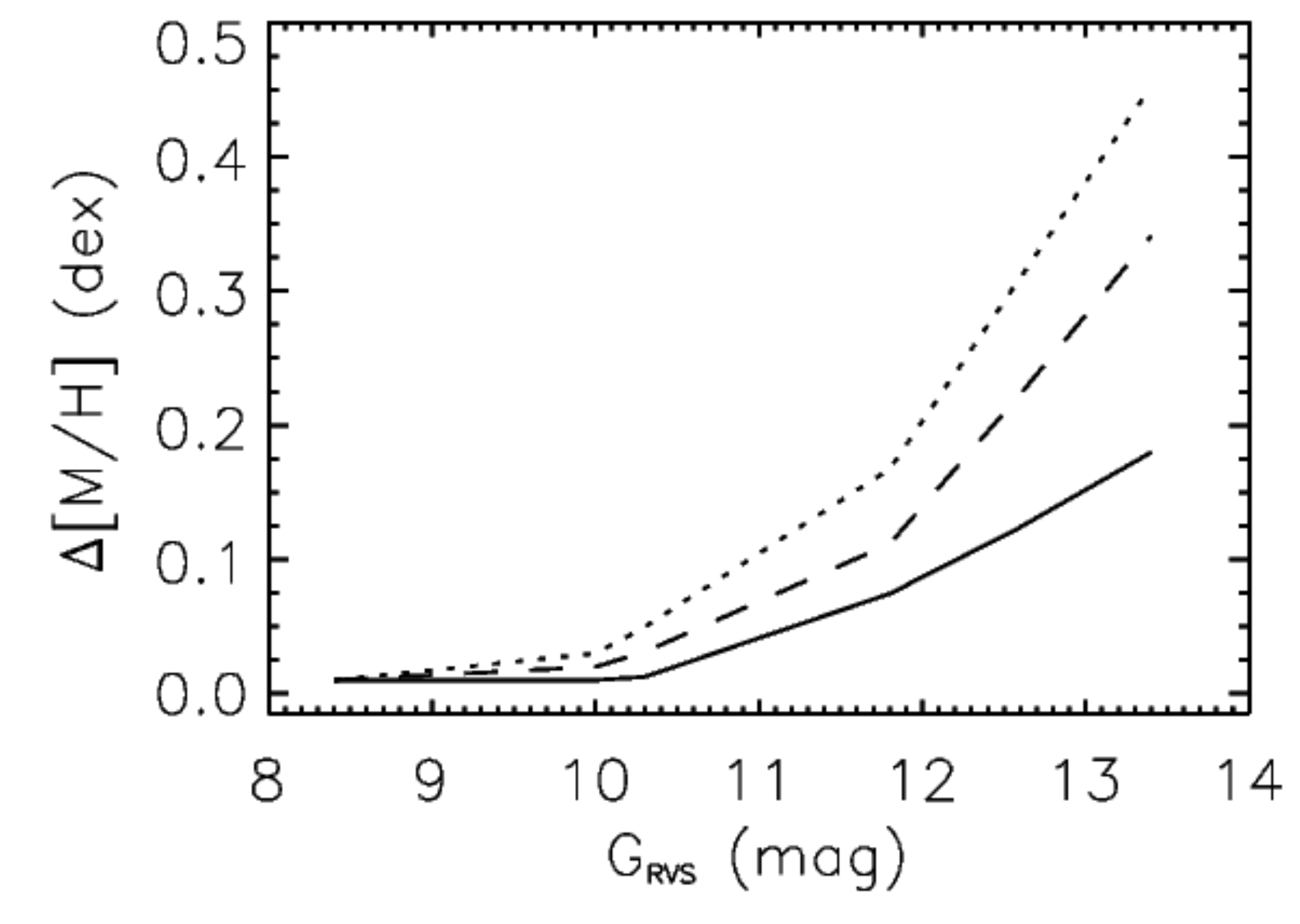}
\caption{Same as Fig.~\ref{FigPerfGSPspecDwarf} but for the A-dwarf
  stars defined in Sect.~\ref{final}.}
\label{FigPerfGSPspecHot}
\end{figure}

Figure~\ref{FigPerfGSPspecHot} shows the expected errors (defined
again as the 68\% quantile of the $\Delta \theta$ distributions) for
A-type stars. As for FGK stars, three different curves are reported
for each stellar parameter, illustrating the behaviour for metal-rich,
metal-intermediate and metal-poor stars.

First of all, we can conclude that the parametrisation of hot stars
with \GRVS $\la$12.5, is expected to be very good (and actually
excellent for the stellar surface gravity). As an example, the typical
error in \meta \ for hot metal-rich stars will be smaller than 0.1~dex
down to that magnitude. In fact, this results from the pressure sensitivity of the Paschen
lines that is a classical luminosity indicator for early-type stars,
specially for \T $\ga$9\,000~K. This comes from the pressure dependence of
the Stark effect. As a consequence, thanks to this important gravity
indicator, hot stars show no dependence of the gravity estimation
accuracy with any atmospheric parameter (c.f. Fig.~\ref{FigErreur}
left middle panel, and Fig.~\ref{FigErreurlogg} bottom panels).  Only
a small degradation with the S/N is detected.
In contrast, both the estimation of the effective temperature and
the metallicity are very sensitive to \T. In fact, the number of
metallic lines drastically decreases for hot stars spectra.

Finally, even for faintest stars, the parameter accuracy is high
enough to allow their classification into the main stellar classes
(spectral subtypes, dwarf/subgiant/giant luminosity classes, with
errors in \g \ lower than approximately 0.2~dex). The stellar
metallicity is expected to be recovered with an error smaller than
0.8~dex for the faintest early-type stars (being smaller than
0.3~dex for A-type metal-rich stars). We recall that A-type dwarfs
are bright stars that will allow to extend the RVS sounded volume
up to distances of 5~Kpc from the Sun (for \GRVS=14).

\subsection{Error correlations and parameter degeneracies}
\label{correlations}

\begin{figure*}[t!]
\includegraphics[height=5.1cm,width=18.cm]{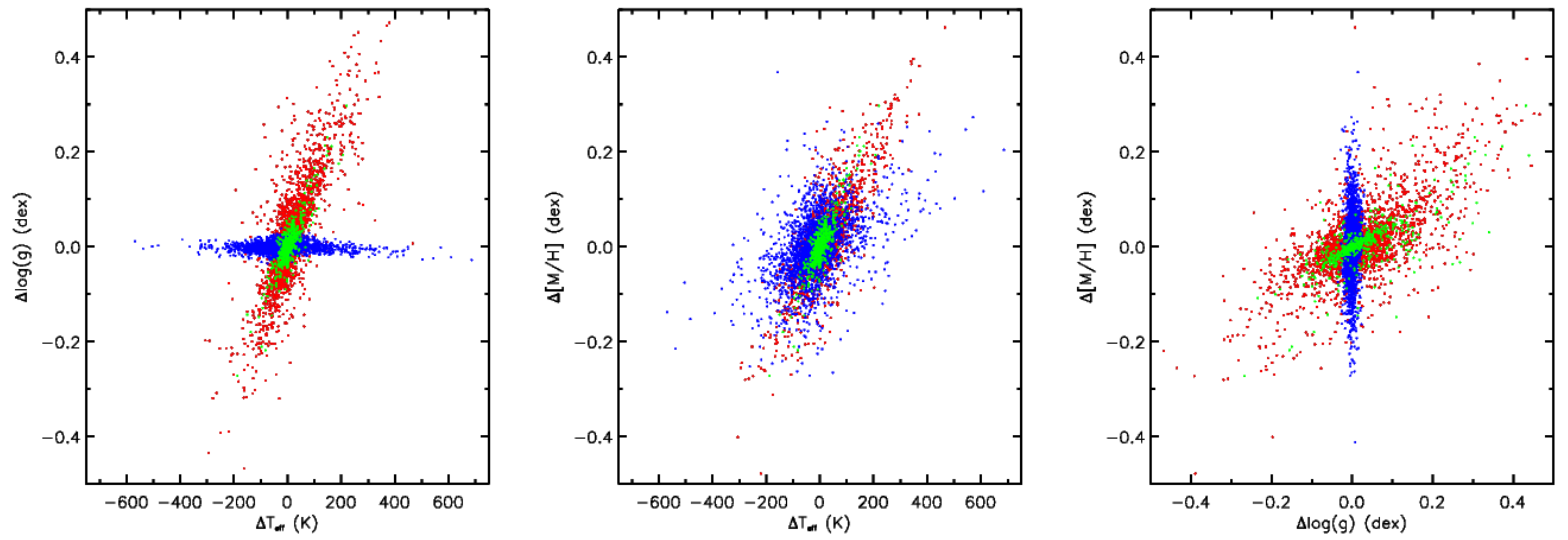}
\caption{Correlations between the residuals of the main atmospheric
  parameters for spectra at \GRVS = 10.3 (S/N$=125$). Cool FGK-dwarf, cool K-giant,
  and hot BA-spectral type stars are plotted in red, green and blue,
  respectively.  The main shape of these correlations does not change
  with \GRVS , only their amplitude varies.}
\label{FigErreurCorrel}
\end{figure*}

One important aspect to be considered in any parametrization exercise
is the existence of error correlations. They not only inform about the
robustness of the results, but also on the possible physical sources
of parameter degeneracies.  Figure~\ref{FigErreurCorrel} shows the
parameter error correlations at a given magnitude, \GRVS = 10.3 (S/N$=125$), chosen
for being the high quality regime, although with a high enough error
amplitude for their analysis. Different colours have been assigned to
FGK-dwarfs (red), K-giants (green) and BA-type stars (blue).  First of
all, a strong correlation is visible between the errors in surface
gravity and effective temperature for the cool FGK-dwarfs. 
This comes from a known degeneracy
between these two parameters \citep{Kordo11a}.  On one hand, the wings
of the CaII lines, carrying much of the information in the RVS domain,
grow proportionally to \g$^{1/3}$ for cool main sequence stars, but
they also strongly depend on the \T.  This implies that differences in
the spectra with rather different parameters are very small, causing
the error correlations seen in Fig.~\ref{FigErreurCorrel}. This
degeneracy is more important in the low metallicity regime, for which
less metallic lines, carrying additional information on \g , remain in 
the spectra. Moreover, as a consequence of the \T-\g \ degeneracy,
the third atmospheric parameter, \meta , is also more difficult to
constrain, showing also error correlations with the other two.

On the other hand, K-giants do not suffer as much as
dwarf stars from this \T-\g \ error correlation, as shown by the
green points of Fig.~\ref{FigErreurCorrel}. This is because, as
already discussed in Sect.~\ref{FGK}, the parameterisation is easier
for giant stars than for dwarfs, with more uncorrelated parameter
variations.

Finally, BA-type stars (blue points in Fig.~\ref{FigErreurCorrel}),
thanks to the presence of a strong and sensitive surface gravity indicator
(the Paschen lines), are not affected by \T-\g \ or \meta-\g \ degeneracies
This is illustrated by the flat behaviour of the \g \ errors as a 
function of the \T \ ones (left panel of Fig.~\ref{FigErreurCorrel}),
and the absence of a relation between the \meta \ errors and the \g \ ones
(right panel). Only a small correlation of the metallicity errors
with the temperature ones (middle panel) seems to exists, in agreement
with the discussions of Sect.~\ref{hot}.

 
\section{Comparison with the expected performances from \gaia spectrophotometric data}
\label{GSPPhot}

In a study rather similar to the present one, \citet{Liu12} reported
the expected performances of stellar parameterisation from \gaia BP/RP
spectrophotometry. \citet{Liu12} analysed the results of different
tested methods within the context of the DPAC/{\it Working Group:
Global Stellar Parametrizer - Photometry} (\gspphot).  Some of the
end-of-mission \gspphot \ expected performances have then been
recently updated in \citet[][see their Tab.~4; see also Andrae et al., 2015, 
in preparation]{BailerJones13}.  The
present section discusses the comparison between the expected
end-of-mission parametrisation performances between RVS (\gsp) and
BP/RP (\gspphot) data for stars brighter than \GRVS $\la$ 15. We
recall that stars fainter than \GRVS$\sim$15 will have only BP/RP
based parameters. Moreover, it should also be pointed out that the present
\gsp \ analysis is performed for simulated RVS spectra that are not affected by
any interstellar extinction, contrary to the \citet{Liu12} results that rely on
BP/RP spectra showing a large range of a priory unknown interstellar extinction.
As a consequence, we will therefore assume in the following discussion that 
\gsp \ is insensitive to interstellar extinction and we will only
consider the \citet{Liu12} results related to the smallest extinctions.
Finally, although  
the tested random samples, the adopted
statistical criteria and, the detailed performances for different types
of stars and magnitudes are not exactly the same in our study and in
the \citet{Liu12} or the \citet{BailerJones13} ones, a rough
quantification of the expected differences can still be performed.  We
point out that, in the following (as in the core of all these papers),
the reported uncertainties in the recovered stellar atmospheric
parameters refer to internal errors only, i.e. relative star-to-star
uncertainties.

For the purpose of this \gspphot \ and \gsp \ comparison, we adopted
the relationship between the $G$-band (\gaia white light) and \GRVS
\ magnitudes already presented in Tab.~\ref{Tab_GRVS} and Fig.~\ref{FigColorColor}.
It is then found
that the \GRVS \ magnitude is brighter than the $G$-band one of about
0.3~mag for A-type stars, of 0.6 to 0.7~mag for F-type stars (with the
magnitude range reporting the metallicity effect from metal-poor to
metal-rich stars), of 0.8 to 1.0~mag for G-type stars and of 1.2 to
1.4~mag for the K spectral type. The variation between cool giants and
cool dwarf stars is weak and has been neglected.

%

%


First, the study of \citet{Liu12} reveals that the \gspphot \ stellar
parametrisation is performed with almost the same efficiency as long
as $G \la 15$ and starts to degrade for fainter stars, only.  This is
also confirmed by the Tab.4 of \cite{BailerJones13} in which the
performances at $G$=9 and $G$=15 are almost identical.  On the
contrary, the \gsp \ parametrization degrades earlier, for \GRVS
fainter than $\sim$11 or 12~mag, depending on the metallicity.  

As pointed out in Sect.\ref{MAR}, the Q68$_\theta$ and rms$_\theta$ statistical indicators can be
  assumed to be almost identical for low S/N RVS spectra.
  This allows us to compare our Tab.~\ref{TabPerfGSPspec}  and the Tab.4 of
\citet[][]{BailerJones13} to roughly deduce performance differences.
 Of course, this
comparison can only show the tendencies suggested
by tests on simulated data, neglecting external errors, mismatches 
between real data and models, parametrization methods optimisation, etc... :

\begin{itemize}
\item Bright dwarf and giant A-type stars (\GRVS $\la$ 12.5) should probably be
  always better parametrized in \T , \g \ and \meta \ from their RVS
  spectra. We recall that very good surface gravities and global
  metallicities (with an accuracy better than 0.1~dex) will be
  available for such type of stars from their \gsp
  \ parametrisation.  For A-stars fainter than \GRVS $\sim$ 12.5,
  although \g \ and \meta \ are still better estimated from their RVS
  spectra, the effective temperature derived from BP/RP data should be
  more accurate.
\item The \gsp \ effective temperature of F-type stars should probably be
  favoured for \GRVS $<$ 12.5. Their \gsp
  \ surface gravity and global metallicity should also be adopted as
  long as \GRVS $\la$ 13. Their accuracy should be better than
  0.1~dex when \GRVS $<$ 12.
\item \gsp \ stellar parameters of GK-spectral type stars should be
  adopted as long as \GRVS $\la$ 12.5.
\item The global \alfaFe \ chemical index will be estimated from the
  \gsp \ pipeline only for FGK-spectral type stars.  Uncertainties of
  the order of 0.1~dex (or even smaller) are expected as long as \GRVS
  $\la$ 12-12.5, depending on the metallicity.
\end{itemize}

In summary, it can be concluded that, for all the considered stellar types,
the stars brighter than \GRVS $\sim$ 12.5 (S/N$=20$) will be very efficiently parametrized
by the \gsp \ pipeline, including good estimations of the \alfaFe \ chemical
index. From these stellar atmospheric parameters, individual chemical
abundances (such as Fe, Ca, Ti, Si,...)  will be derived with an
expected uncertainty smaller than 0.1~dex for most of the RVS sample with 
about \GRVS $\la$ 12 (S/N$\ga 35$), i.e. for a few million of targets.  For faintest stars
that are better parametrised from their BP/RP photometry, a \T \ input
from \gspphot \ as an initial condition for \gsp \ will allow the 
improvement of its final \g , \meta \ and \alfaFe \ estimates.  Such a
\gspphot / \gsp \ link is already implemented in the Apsis processing
system developed by the CU8 and combined performance tests are being
implemented.

Finally, we also stress that the spectral parametrization for
extincted stars should be easier with the \gsp \ pipeline since a
\T-extinction degeneracy appears in the parametrisation of BP/RP
low-resolution spectra with too large line-of-sight interstellar
extinction (assuming that their brightness in the RVS band is not too
faint to collect high enough S/N spectra). In those cases,
a feedback from \gsp \ to \gspphot, in a second iteration of the
analysis cycle, will also improve the final parameters estimations.

\section{Comparison with other spectroscopic surveys}
\label{Surveys}
A suite of ground-based vast stellar spectroscopic surveys mapping the Milky Way
is revolutionizing the observational information about Galactic stellar populations.
Their synergy with the Gaia mission relays not only on the sounded spatial volume,
but also on their spectral resolution and covered wavelength domain. These two
characteristics primarily determine their corresponding performances in the stellar parameters 
and chemical abundances estimation.

The Sloan Digital Sky Survey project, in its series of operations (SDSS I, II and
III) has published about 250 000 spectra (R=1800) from the Sloan Extension for Galactic Understanding and
Exploration \citep[SEGUE;][]{YannyRockosi09}. SDSS spectra have provided only limited information on
the structures revealed in the SDSS photometry, but they produced \T \ and \g \ estimates to 250~K and 0.5~\gunits \ respectively, 
and [Fe/H] abundances to 0.3~dex for stars with 14$<$r$<$19~mag \citep[][]{Schlesinger12}. SEGUE data 
overlap the RVS targets only in the RVS fainter magnitudes domain. Due to the larger wavelength coverage
of the SEGUE data, its \T \ estimations are generally better than the expected \gsp  \ ones. However, the
higher RVS resolution should allow more precise measurements of \g, and \meta.

The RAdial Velocity Experiment \citep[RAVE;][]{Steinmetz06}  is obtaining accurate radial velocities ($<$ 5
km/s) and global metallicities for 5$\cdot$10$^5$ stars with J$<$12 from spectra with R$=$7500. RAVE is also estimating
the individual abundances of some elements for several thousand stars. This project, due to its rather bright
magnitude limit, corresponding to about  \GRVS$=$12 for solar type stars, is probing essentially the Galactic 
discs populations. In terms of parameter estimations, the RAVE internal errors at SNR$=10$ 
(about \GRVS$=$12) are of  350~K for \T, 0.5~dex \ for \g \ and 0.3~dex for \meta \  
for solar-type stars \citep[cf.][Table~1]{Kordo13}. More generally, the \gsp \ performances should always be better than the
RAVE ones, as expected from the RVS fainter magnitude limit and higher resolution.

More recently, the Large sky Area Multi-Object fiber Spectroscopic Telescope \citep[LAMOST;][]{Zhao12} project has
implemented a survey dedicated to Galactic exploration \citep[LEGUE;][]{Deng12}. The LEGUE survey plan includes spectra 
for 2.5 million stars brighter than $r<19$ and an additional 5 million stars brighter than $r < 17$. The  magnitude distribution 
depends on the telescope throughput and the survey resolution, much lower than the RVS one, is R$=$1800. 
\cite{Xiang15} estimate the uncertainties of the LAMOST stellar parameter pipeline to be of about 150~K in \T, 0.25~dex in \g \
and 0.15~dex in [Fe/H]. On the other hand, for red giant stars, \cite{Liu14}  report typical errors in metallicity in the range 0.15 to 0.30~dex.
On the other hand, similarly to the SEGUE survey, LEGUE data mainly overlap the RVS observations in the faint 
magnitudes domain, for which BP/RP data will also be available. 

The results of the first low-resolution surveys revealed the key role of the stars chemical information to
disentangle the Milky Way stellar population puzzle, motivating a new era of ground based high-resolution spectroscopic 
surveys. Three of them will be active from the ground during the period 2015-2019:
The Gaia-ESO Survey \citep[GES;][]{Gilmore12},  the SDSS Apache Point Observatory Galactic Evolution 
Experiment \citep[APOGEE;][]{Eisenstein11}  and the Galactic Archaeology with AAO HERMES 
\citep[GALAH;][]{Zucker12} survey. All these surveys, thanks to their larger wavelength coverage
and resolution will provide more accurate parameters than Gaia/RVS for a subsample of stars.
Nevertheless, only the GALAH survey, targeting about one million stars with  V$<$14, is expected to 
have an important overlap with the RVS. This overlap will correspond, in any case, to only less than one tenth 
of the RVS targets with \GRVS$<$13. The GES survey is mainly targeting faint stars (14$<$V$<$19)
thanks to the Very Large Telescope FLAMES/GIRAFFE facility ((R$\sim$20\,000) and it will primarily 
complement the Gaia/BPRP parameter estimations. In the RVS magnitude domain, only a small GES sample
of 10$^4$ G-stars within 2 kpc of the Sun (12$<$V$<$14.5, corresponding to about 11$<$\GRVS$<$13.5)
is being observed with the FLAMES/UVES spectrograph (R=40 000). Finally, the APOGEE survey,
is preferentially targetting high extinction regions of the disc and the bulge in the range 8$<$H$<$13.8.
Although the magnitude coverage overlaps the RVS one, the APOGEE targeted fields are
characterized by a high stellar crowding that limits the RVS observations. Therefore, APOGEE will
mostly complement the RVS survey near the Galactic plane, rather than overlapping it.

In conclusion, the RVS based stellar parameters will provide precious information about the Galactic populations
in the bright part of the Gaia volume, for a number of stars tens of times higher than what will
be provided by currently on-going and planned spectroscopic surveys from the ground. Those surveys, specially
those at high spectral resolution, will nevertheless be crucial for the Gaia/RVS parameters validation and to
complement them with precise chemical abundances for a subsample of stars.

\section{Conclusion}
\label{Conclu}

In this work, after having analysed the results of different
independent methods, we have estimated the end-of-mission expected
parametrization performances of the \gaia \ DPAC pipeline (\gsp) in
charge of the RVS stellar spectra atmospheric parameters and chemical
abundances derivation.  The estimated accuracies, as a function of
stellar types and magnitudes are summarized in
Tab.~\ref{TabPerfGSPspec} and in Fig~\ref{FigPerfGSPspecDwarf} to
\ref{FigPerfGSPspecHot}.

The reported uncertainties in the recovered stellar atmospheric
parameters refer to internal errors only, i.e.  relative star-to-star
uncertainties. Total errors will be, in many cases, dominated by
external ones (partly caused by the possible synthetic spectra
mismatches with respect to real observed ones) and they will be estimated from the
analysis of real Gaia RVS spectra of benchmark reference stars, during a results validation
phase. Nevertheless, the internal errors reported here permit to
clearly identify, and quantify in detail, the enormous variety of
science cases that will be obtained from the interpretation of pure
\gaia \ data (without any need of references to external catalogues).

The \gsp \ pipeline will be optimised in the light of the first analysed 
real RVS spectra over the next  year. Increasingly improved
versions of the  Apsis 
\gsp \ module are delivered at each operations cycle. \gsp \ is expected to 
 be running in operations cycle~4 in 2017, with a possible contribution from 
 the third \gaia data release.
The current \gsp \ version, integrated in the general Apsis chain,
already meets the tight requirements in processing speed (17~Mflops per
source), needed to repeately treat tens of million of spectra.

Our tests, including first estimations of the impact caused by
the on board detected stray light contamination, show that the contribution of the RVS based stellar parameters will be unique
for stars with \GRVS$\la$12.5 (a few tens of million of stars).  On the
one hand, the \gsp \ parameters will probably be more accurate than the majority
of the parameters derived from the spectrophotometry in that
magnitude range.  This will allow, thanks to the use of the \gaia
parallaxes, a better estimation of the stellar evolution phase and, as
a consequence, of the isochrone based age estimations (for which the
effective temperature accuracy is a dominant source of
error). Accurate stellar ages will be one of the revolutions in Milky
Way astrophysics that the \gaia mission will accomplish, and the RVS
data will strongly contribute to it, sharpening our view of the
Galactic history in a volume of very precise measurements (up to $\sim$8~kpc
from the Sun for K-giants and $\sim$1~kpc for G-dwarfs).

On the other hand, accurate metallicity and chemical abundance
measurements as the \alfaFe \ content are today recognized as crucial
information for the understanding of the highly complex evolution of
Galactic stellar populations. As an example, the classical kinematically-based
definitions of the thin and the thick disc populations blurred our comprehension 
of the Galactic disc substructure \citep[c.f.][]{Bovy2012, Recio-Blanco14}.
The RVS chemical abundance estimations, with accuracies better than 0.1~dex,
will therefore be a unique and precious sample of several million of stars 
for which many pieces of the Milky Way history puzzle will be available, 
with unprecedented precision and statistical relevance.


\begin{acknowledgements}
We thank the Centre National d'Etudes Spatiales (CNES, France) and the
French CNRS/INSU for continuous support for the preparation of the
\gaia mission.  This work benefited from travel supports from the
European Science Foundation through the GREAT Research Network
Program.  Part of the computations have been done on the 'Mesocentre
SIGAMM' machine, hosted by the Observatoire de la C\^ote d'Azur.
The first two authors would like to thank
Naia for her (too) numerous (and efficient) attempts to postpone the revision of this paper.
We warmly thank C.A.L. Bailer-Jones for his constructive remarks that helped
to improve this article. We are also sincerely grateful to D. Katz for his help concerning
the in-flight RVS characteristics.
\end{acknowledgements}

\bibliographystyle{aa.bst}
\bibliography{GSPSpec}

\end{document}